\documentclass[10pt,aps,prd,twocolumn,showpacs,superscriptaddress,nofootinbib,nobibnotes,longbibliography,floatfix]{revtex4-2}

\usepackage[utf8]{inputenc}
\usepackage[T1]{fontenc}
\usepackage{bm}
\usepackage{mathtools,amsmath,amssymb,amsfonts,mathrsfs,eucal,graphicx,tensor,csquotes,accents,commath,chngcntr,siunitx}
\usepackage[dvipsnames]{xcolor}
\usepackage[unicode]{hyperref}
\hypersetup{colorlinks=true, citecolor=MidnightBlue,
            linkcolor=MidnightBlue, urlcolor=MidnightBlue, linktocpage=true}
\usepackage[normalem]{ulem}
\usepackage{orcidlink}

\newcommand{\orcid}[1]{\href{https://orcid.org/#1}{\includegraphics[width=10pt]{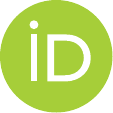}}}

\begin{document}

\author{Saulo Albuquerque \orcid{0000-0003-2911-9358}}
\affiliation{Dipartimento di Scienze Pure e Applicate, Universit\`a degli Studi di Urbino ``Carlo Bo'', Urbino, Italy}
\affiliation{Departamento de F\'isica, Universidade Federal da Para\'iba, Caixa Postal 5008, Jo\~ao Pessoa 58059-900, PB, Brazil}
\email{saulo.soaresdealbuquerquefilho@uniurb.it}

\author{Sebastian H. V\"olkel \orcid{0000-0002-9432-7690}}
\affiliation{Max Planck Institute for Gravitational Physics (Albert Einstein Institute), D-14476 Potsdam, Germany}
\email{sebastian.voelkel@aei.mpg.de}

\date{\today}

\title{Bayesian analysis of analog gravity systems with the Rezzolla-Zhidenko metric}

\begin{abstract}
Analog gravity systems have the unique opportunity to probe theoretical aspects of black hole physics in a controlled laboratory environment that one cannot easily observe for astrophysical black holes. 
In this work, we address the question of whether one could use controlled initial perturbations to excite the black hole ringdown and infer the effective black hole metric. 
Using a theory-agnostic ansatz for the effective metric described by the Rezzolla-Zhidenko metric and evolving perturbations on that background, we quantify with Bayesian analysis what regions of the effective spacetime could be constrained in experiments. 
In contrast to standard ringdown analyses based on quasinormal mode extraction, a laboratory-controlled setup, in combination with our framework, allows one to model the entire signal, including the prompt response and possible effects of late-time tails. 
Therefore, it has the intriguing advantage of not relying on start and end times when the superposition of quasinormal modes is a good signal approximation. 
It also avoids the nontrivial question of how many modes are present.  
We demonstrate that this approach is feasible in principle and discuss opportunities beyond this study. 
\end{abstract}

\maketitle

\section{Introduction}\label{intro}

Black holes are among the most extreme objects in our Universe. 
Theoretical considerations trace back a long time in the history of physics, i.e., to the Gedankenexperiment that objects might have a gravitational field so strong that not even light could escape~\cite{Michell:1784xqa,1799AllGE...4....1L}. 
Even within Einstein's general theory of relativity~\cite{Einstein:1916vd}, it took some time from Schwarzschild's solution describing a nonrotating black hole~\cite{Schwarzschild:1916uq}, to the rotating generalizations known as Kerr metric~\cite{Kerr:1963ud}, and Kerr-Newman metric~\cite{Newman:1965my}. 
With the ongoing efforts of the direct measurement of gravitational waves from binary black hole systems by the LIGO-Virgo-KAGRA Collaboration~\cite{LIGOScientific:2016aoc}, the imaging of supermassive black holes by the Event Horizon Telescope Collaboration~\cite{EventHorizonTelescope:2019dse,EventHorizonTelescope:2022wkp}, and the measurements of S-stars orbiting Sgr A$^*$~\cite{GRAVITY:2020gka}, black holes are a central player in modern physics.  

Already half a century ago, works by Hawking, Bekenstein, and others opened a more theoretical field in black hole physics related to quantum field theory in curved spacetimes and discovered black hole thermodynamics~\cite{Bekenstein:1973ur,Bardeen:1973gs,Hawking:1975vcx}. 
Although related predictions such as Hawking radiation cannot be easily verified with astrophysical black holes, they could, in principle, provide us with a gateway to exploring the unsolved problem of quantum gravity. 

Somewhere in between the study of real black holes and conducting feasible laboratory experiments lies the field of analog gravity~\cite{Barcelo:2005fc}. 
Pioneered by Unruh~\cite{Unruh:1980cg}, there is a wide range of systems whose properties can be related to those of gravitational systems. 
The analogy on a formal level, i.e., the mathematical equivalence of equations, can then be used to explore black hole effects in laboratory systems~\cite{visser1993acousticpropagationfluidsunexpected,bec1,bec2,Novello2002,Volovik2003, Nakano:2004ha, Unruh2007, Vieira:2021xqw, Steinhauer:2014dra,PhysRevLett.117.271101,Barcelo:2018ynq,2022PhRvD.105d5015V,kostashoracio,solidoro2024quasinormalmodessemiopensystems,DelPorro:2024tuw,keshet2024ringdownhawkingradiationconnectionreal,Liao:2018avv,Moreno-Ruiz:2021qrf,Haller:2022pfj,smaniotto2025blackholespectroscopygiantquantum}. 
Early ideas focused on hydrodynamical systems like a draining vortex, where the sound horizon plays the role of the black hole horizon. 
Since then, more recent developments have paved the way to study quantum matter like Bose-Einstein condensates or liquid helium, thus reaching out to completely different areas of modern physics~\cite{PhysRevLett.123.161302,Steinhauer:2015saa,Clovecko:2018qnj,Braunstein:2023jpo,Svancara:2023yrf,extra1,extra2,extra3,extra4}. 
In such experiments, a variety of observations had been made, including the analog of Hawking radiation~\cite{hawkingbec1,hawkingbec2,hawkingbec3,Rousseaux:2007is, PhysRevLett.117.121301, 2019Natur.569..688M, 2021NatPh..17..362K}, superradiance~\cite{2017NatPh..13..833T}, and quasinormal modes~\cite{Torres:2020tzs}. 

This work outlines a framework that explores the possibility of inferring an analog system's effective black hole metric by analyzing time-domain wave propagation signals. 
More specifically, we first adjust the Rezzolla-Zhidenko (RZ) metric as a flexible model for analog black holes without rotation~\cite{Rezzolla:2014mua}; see Ref.~\cite{Konoplya:2016jvv} for the rotating case. 
We then study the scalar wave equation on this background as a prototype for surface-wave propagation in a draining vortex~\cite{Visser:2004zs,Patrick:2018orp,2017NatPh..13..833T}. 
Since an experimental setup allows one, in principle, to fine-tune the initial conditions for wave propagation, we can study a case that would only be of academic interest in the case of astrophysical black holes. 
By preparing the initial data, we can model the analog system's full time evolution solely using the free parameters of the RZ metric. 
By utilizing this modeling ansatz in a Bayesian Markov chain Monte Carlo (MCMC) code, we can explicitly infer the effective background metric even in the presence of noisy data. 

We focus on Gaussian initial data sent toward the analog black hole, which then gets partially reflected by an effective potential barrier. 
As measured by an observer far from the vortex, the analog model's response will qualitatively correspond to the typical ringdown of astrophysical black holes emerging from binary black hole mergers. 
In such an application, one knows that a sum of quasinormal modes will be a good approximation of the signal at intermediate times after nonlinear effects from the merger disappear and before late-time, power-law tails dominate~\cite{Price:1971fb,Leaver:1986gd,Buonanno:2006ui}. 
Finding the starting and end times when the quasinormal mode approximation is valid, as well as choosing the number of included quasinormal modes, especially for overtones, is highly nontrivial. 
There is no universal recipe to address these issues; e.g., see Refs.~\cite{Baibhav:2023clw,Nee:2023osy,thomopoulos2025ringdownspectroscopyphenomenologicallymodified} for some recent critical works tackling this problem for gravitational waves. 

One can easily circumvent these two significant problems in the analog gravity context. 
Our modeling does not rely on specifying a given number of quasinormal modes or choosing a time window. 
Instead, we can model and analyze the entire time signal. 
Thus, for some initial data, we can model the complete response of any observer, which is qualitatively more similar to an inspiral, merger, and ringdown analysis in the gravitational wave case. 
To be computationally feasible in parameter estimation, the gravitational wave analysis requires waveform models that combine multiple techniques, e.g., traditionally by incorporating post-Newtonian, numerical relativity, and perturbative computations; see Refs.~\cite{Buonanno:1998gg,Ajith:2009bn,Field:2013cfa,Purrer:2014fza} for some pioneering contributions and Refs.~\cite{Maggiore:2007ulw,Maggiore:2018sht,chatziioannou2024compactbinarycoalescencesgravitationalwave} for more comprehensive material. 
In our application, the numerical cost of modeling the analog gravity system is much cheaper, which allows us to carry out the time evolution of the initial data for different RZ parameters during sampling, thus avoiding any possible challenges related to building a robust waveform model first. 

Our main results are that direct numerical modeling during MCMC sampling is practically feasible and that reconstructing the injected analog black hole metric parameters and perturbation potentials is possible. 
By exploring different signal-to-noise ratios (SNRs) and observer locations, we also provide qualitative requirements for possible experiments. 

The rest of this work is structured as follows. 
In Sec.~\ref{methods}, we outline our methods related to the background metric, time evolution, and Bayesian analysis. 
In Sec.\ref{app_results}, we show our application and results and provide further discussion in Sec.~\ref{discussion}. 
Our conclusions can be found in Sec.~\ref{conclusions}. 
We use units in which $G=c=1$.

\section{Methods}\label{methods}

In Sec.~\ref{method}, we introduce the RZ metric; in Sec.~\ref{method2}, we show how it can be related to analog gravity systems; in Sec.~\ref{method3}, we explain our numerical time evolution scheme; and in Sec.~\ref{method4} we outline the Bayesian parameter estimation including our signal analysis.

\subsection{Rezzolla-Zhidenko metric}\label{method}

There are diverse motivations for developing black hole metrics that describe deviations from the Schwarzschild or Kerr solutions. 
It can range from performing theory-agnostic tests of general relativity to approximating numerical solutions of modified black hole spacetimes with analytic functions. 
The literature is quite comprehensive, and we cannot review and discuss all of them. 
The interested reader can find some popular ones in Refs.~\cite{Johannsen:2011dh,Johannsen:2013szh,Rezzolla:2014mua,Konoplya:2016jvv}.

\subsubsection{Rezzolla-Zhidenko metric in $3+1$ dimensions}

In this work, we study a slight modification of the RZ metric, which was originally introduced in Ref.~\cite{Rezzolla:2014mua} for spherically symmetric 3+1 dimensional black holes. 
It was extended as the Konoplya-Rezzolla-Zhidenko (KRZ) metric to axial symmetric black holes in Ref.~\cite{Konoplya:2016jvv}, and for a $D$-dimensional black holes, where $D\geq4$ in Ref.~\cite{Konoplya:2020kqb}. 
The advantage of the (K)RZ metric is its continued fractions expansion, making it very efficient in modeling significant deviations with only a few parameters; see Ref.~\cite{Konoplya:2020hyk} for a demonstration.

We refer to the original work~\cite{Rezzolla:2014mua} for a complete discussion of the RZ metric. 
In the following, we only report the main equations for a general $D$ dimensional case, but we will adjust the angular line element and the radial compactification later in our application. 
The RZ line element is given by
\begin{align}\label{metric}
\text{d}s^2=-N^2(r)\text{d}t^2+\frac{B^2(r)}{N^2(r)}\text{d}r^2+r^2 \text{d}\Omega^2\,,
\end{align}
where the horizon is located at $N(r_0)=0$. 
After introducing a radial compactification $x(r)$ defined by the condition that the horizon is located at $x(r\rightarrow r_0)=0$ and spatial infinity at $x(r\rightarrow\infty)=1$, the metric is written as 
\begin{align}
N^2=x A(x)\,,
\end{align}
where $A(x) >0$ for $0\leq x \leq 1$. 
The functions $A(x)$ and $B(x)$ are now written in terms of 
\begin{align}
A(x)=&1-\epsilon (1-x)+(a_0-\epsilon)(1-x)^2
\\
\nonumber
&+{\tilde A}(x)(1-x)^3,
\\
B(x)=&1+b_0(1-x)+{\tilde B}(x)(1-x)^2,
\end{align}
where 
\begin{align}
{\tilde A}(x)=\frac{a_1}{\displaystyle 1+\frac{\displaystyle
    a_2x}{\displaystyle 1+\frac{\displaystyle a_3x}{\displaystyle
      1+\ldots}}}\,,
\\
\nonumber 
\\
{\tilde B}(x)=\frac{b_1}{\displaystyle 1+\frac{\displaystyle
    b_2x}{\displaystyle 1+\frac{\displaystyle b_3x}{\displaystyle
      1+\ldots}}}\,.
\end{align}
Note that all higher-order coefficients $a_i$ and $b_i$ do not change the metric if one of the previous ones vanishes (for $i > 1$).

\subsubsection{Rezzolla-Zhidenko metric in $2+1$ dimensions}\label{method1}

In $3+1$ dimensions, the radial compactification, as originally proposed ~\cite{Rezzolla:2014mua}, is defined as
\begin{align}
x \equiv 1-\frac{r_0}{r}\,. 
\end{align}
Demanding an asymptotic matching to the Schwarzschild metric one finds
\begin{align}\label{usualcompact}
\epsilon=\frac{2M-r_0}{r_0} = - \left(1 - \frac{2M}{r_0}\right)\,,
\end{align}
which relates ADM mass $M$, horizon location $r_0$ and $\epsilon$ with each other. 

In our applications, however, we want to extend the RZ metric to a $2+1$ dimensional, effective spacetime describing an analog black hole. 
In such a case, one cannot use the original version.  
By taking a careful look at the effective metric for analog models and comparing its functional form with the one of Schwarzschild-like higher-dimensional black hole spacetimes \cite{kostashoracio,Konoplya:2020kqb}, we find that a better choice for the radial compactification coordinate is given (from here on) as
\begin{align}\label{amcompact}
x \equiv 1-\left(\frac{r_0}{r}\right)^{2}\,.
\end{align}

\subsection{Perturbation equations for analog black holes}\label{method2}

Using the RZ metric as a starting point to model analog gravity systems, we assume that the underlying wave equation of a scalar field is given by
\begin{align}
\square_{g} \Phi(t,r,\theta,\phi) = 0\,,
\end{align}
where $\square_{g}$ is the D'Alembertian in the coordinate system of the effective metric $g$. This metric that can either be derived from the specific analog system or be modeled agnostically, e.g., using the RZ metric. 
We assume this structure is valid for classes of analog systems, e.g., a draining vortex~\cite{Schutzhold:2002rf,Dolan:2011ti}. We choose our units such that the speed of sound is $c=1$.

The wave equation in the effective space-time can be simplified to
\begin{align}\label{eq:wave}
\left[ \partial^2_t - \partial^2_{r^*} + V_m(r) \right] \Psi(t,r) = 0\,,
\end{align}
where one uses a radial rescaling and an angular decomposition for cylindrical
coordinates, namely
\begin{align}
\Phi = \Psi e^{im\phi}/r^{1/2}\,.
\end{align}
The new radial coordinate, usually called tortoise coordinate $r^*$,  is defined by
\begin{align}
\frac{\text{d}r^*}{\text{d}r}=\frac{B(r)}{N^2(r)}\,.
\end{align}

For our analog model in $2+1$ dimensions, the radial equation leads us to the following effective potential \cite{extra1, Torres:2022bto}
\begin{align}\label{potam}
V_{m}(r) =& N^2(r)\frac{m^2}{r^2}  -\frac{1}{4 r^2}  \left(\frac{N^4(r)}{B^2(r)}\right) \nonumber \\ & +\frac{1}{4r}\frac{d}{dr}\left[\frac{N^4(r)}{B^2(r)}\right]\,.
\end{align}

If one assumes $g_{tt} =-1/g_{rr}$, which is equivalent to taking $B(r)=1$ in Eq.~\eqref{metric}, the potential simplifies to
\begin{align}\label{potamsym}
V_{m}(r) = f(r)\left(\frac{m^2}{r^2}  -\frac{f(r)}{4 r^2}+\frac{f'(r)}{2r}\right)\,,
\end{align}
where
\begin{align}
f(r)=N^2(r)\,.
\end{align}
Equation ~\eqref{potamsym} reproduces Eq. (78) from Ref.~\cite{extra1} for acoustic perturbations on the radial vortex.

\subsection{Time-domain integration}\label{method3}

The central piece of our modeling is the time evolution of given initial data. 
We implement the widely used staggered leapfrog algorithm, e.g., see Ref.~\cite{Nee:2023osy}, which solves the wave equation Eq.~\eqref{eq:wave} numerically via a finite difference scheme central in time and space
\begin{multline}\label{timeintegration}
\Psi_{j}^{i} = 2\Psi_{j}^{i-1}-\Psi_{j}^{i-2}+\frac{\Delta t^2}{\Delta {r^*}^2}\left(\psi^{i-1}_{j+1}-2\psi^{i-1}_{j}+\psi^{i-1}_{j-1}\right)\\-\Delta t^{2}\,\Psi_{j}^{i-1}V_{j}\,.
\end{multline}
Here we define $\Psi^{i}_{j} = \Psi(t_{i}, r^*_{j})$, $V_{j} = V_m(r^*_{j})$ and $\Delta t$ and $\Delta r^*$ define the temporal and spatial resolution as follows. 
We choose a spatial grid of 2000 equally distant points in the tortoise coordinates between $r^*=-100$ and $r^*=200$, which gives us $\Delta r^*=300/2000=0.15$. The time resolution is defined by $\Delta t= 1/4 \Delta r^{*}$.
We discuss the impact of this resolution in Sec.~\ref{disc_biases}.

We model our initial data as an ingoing Gaussian wave packet
\begin{align}\label{initialdata}
\Psi(t,r^*)\big|_{t=t_\mathrm{start}} &= Ae^{-\frac{(r^* - r^*_0)^{2}}{2d^{2}}},
\\
\frac{\partial}{\partial t} \Psi(t,r^*)\big|_{t=t_\mathrm{start}} &= \frac{\partial}{\partial{r^*}} \Psi(t,r^*)\big|_{t=t_\mathrm{start}}\,,
\end{align}
which in $t=t_\mathrm{start}$ is centered at $r^*_0=50$ and has an initial width of $d=2$, initial amplitude $A=4$, and initial time $t_\mathrm{start}=0$. The pulse evolves toward the potential barrier with a speed of sound set to $c=1$.
We choose the potential by fixing the angular momentum as $m=3$. 
As input for our Bayesian analysis in Sec.~\ref{method4}, we use the time series $\Psi(t, r^*=r^*_\text{obs})$, where $r^*_\text{obs}$ is the location of an observer.

\subsection{Bayesian analysis}\label{method4}

Bayesian analysis is a popular tool to study the inverse problem when one analyzes noisy data $\bm{d}$ to infer the probability distribution of the parameters $\bm \theta$ describing a given model. 
Bayes' theorem connects data and model via
\begin{align}
p\left(\bm{\theta} | \bm{d} \right) = \frac{ p \left(\bm{d} | \bm{\theta} \right) p \left(\bm{\theta} \right)}{p \left( \bm{d} \right)}\,,
\end{align}
where $p\left(\bm{\theta} | \bm{d} \right)$ is the posterior distribution, which describes the probability of the parameters given the data. 
Bayes' theorem says it is equal to computing the likelihood $p \left(\bm{d} | \bm{\theta} \right)$, which describes the probability of the data given the parameters times the prior $p\left(\bm{\theta} \right)$, which is the probability of the parameters before analyzing data, divided by the evidence $p\left( \bm{d} \right)$, which is the probability of the data itself. 
The evidence is important for model comparison, however, it is not needed for MCMC sampling. See Ref.~\cite{sivia2006data} for a pedagogical introduction to the method.

Calculating the posterior distribution analytically can only be done in special cases. 
A powerful approach to numerically compute the posterior distribution is to draw samples from it using MCMC techniques. 
Standard approaches, like the Metropolis-Hastings algorithm~\cite{Metropolis:1953am}, only require the knowledge of the likelihood and the prior, but not the evidence. 
During MCMC sampling, posterior samples are drawn by computing the ratio of likelihood times prior of two consecutive steps. 
This can become computationally expensive because the likelihood and prior have to be evaluated many times to have a large enough sample size to approximate the posterior distribution, especially in higher-dimensional parameter spaces with multimodal distributions. 
To perform the MCMC analysis, we utilize the \texttt{emcee} sampler, which has been introduced in Ref.~\cite{Foreman-Mackey:2012any}. 
It is based on the affine-invariant ensemble sampler proposed in Ref.~\cite{2010CAMCS...5...65G}, which makes it more efficient than the standard Metropolis-Hastings algorithm. 
In all of our MCMC analyses in Sec.~\ref{app_results}, we have combined 2 independent \texttt{emcee} runs with 50 walkers with 1000 steps each. 
In total, this corresponds to 100000 draws from the posterior distribution. 

In this work, the free parameters are those considered for the RZ metric, and the data will be described by the ringdown signal of an observer far away from the effective black hole. 
Our prior is described by a uniform distribution for each RZ parameter within a range reported in the respective corner plots of their final MCMC results, in Sec.~\ref{app_results}. 
When using multiple RZ parameters at once, it is not trivial to constrain the full RZ parameter space to only contain physical solutions for the metric; see Refs.~\cite{Kocherlakota:2022jnz,kocherlakota2022commentanalyticalboundsrezzollazhidenko}. 
In this work, we assume that $g_{tt}$ and $g_{rr}$ are not changing sign. 
This ensures that $r_0$, as used to define the compactified coordinate $x$, describes the outermost horizon. 
We enforce this condition by rejecting any proposal violating it during sampling. 
While the physical justification for a specific condition might be arguable depending on the analog models one is interested in, this work mainly guarantees that the time evolution is well-defined.

In the following, we outline a basic signal-processing treatment motivated by gravitational wave research but with a few simplifications. 
As a proof of principle, we consider a Gaussian likelihood in the time domain
\begin{align}\label{likelihood}
p(\bm{d}|\bm{\theta}) 
&\propto 
\exp\left(- \frac{1}{2} \left< \bm{d} -m(\bm{\theta}) | \bm{d}-m(\bm{\theta}) \right> \right) \,,
\end{align}
where $\bm{d}$ is the observed time series and $m(\bm{\theta})$ is our model. 
In Sec.~\ref{app_results}, we will set $\bm{\theta}$ to be the nonzero parameter of the RZ metric ($r_0, a_0, b_0, \epsilon$) 
and $\bm{d}$ will be given by the theoretically measured time-series $\Psi(t,r^*)$ evaluated at specific locations $r^*$ and for a certain time span $t \in[t_\text{start}, t_\text{end}]$. 
The model $m(\bm{\theta})$ is the time series for a specific $\bm{\theta}$ using the same initial data Eq.~\eqref{initialdata} in the numerical time evolution code from Eq.~\eqref{timeintegration}. In this framework, we assume that both the sampling signals $m(\bm \theta)$ and the injections $\bm d$ are modeled equivalently by Eqs.~\eqref{timeintegration}  and \eqref{initialdata}. 
Hence, it can be stated that the injected signal is equivalent to $\bm d = m(\bm{\theta^\mathrm{inj}})$ .

The inner product between two generic time series $h(t)$ and $g(t)$ is defined as
\begin{align}\label{eq_inner}
 \left< h | g \right> &=\int_{t_\mathrm{start}}^{t_\mathrm{end}} \frac{h(t) g(t)}{S} \text{d}t\,,
\end{align}
where we assume that $h,g$ are real-valued functions of time and $S$ will be a time-independent constant. We set 
$t_{\text{start}}$ as the initial time when the Gaussian pulse is generated,  i.e. $t_{\text{start}}=0$. The endpoint, $t_\text{end}$, is chosen to be sufficiently large to ensure that the scattered signal reaches the observer, allowing for the detection of all ringdown damped oscillations before the signal vanishes and becomes negligible. 

Our statistical treatment in Eqs.~\eqref{likelihood} and \eqref{eq_inner} corresponds to white noise, and we leave experiment-specific noise modeling for future studies because those depend on the underlying experiment. 
Typically, this would require one to work in the frequency domain, and $S(f)$ would describe the noise properties. 
The SNR $\rho$ of a given time series $h$, e.g., $\Psi(t)$, is defined as
\begin{align}\label{eq_SNR}
\rho^2 = \left< h  | h \right>\,,
\end{align}
and in order to
perform MCMC sampling at a specific SNR for a given simulation, one must choose $S$ accordingly. 

Finally, we conclude the review of our statistical methods by noting that we consider noiseless injections because we are interested in a clean injection/recovery study. 
This means we do not explicitly add specific Gaussian noise realizations to the time series. 
The impact of Gaussian noise of different strength on the uncertainty region of inferred parameters is  controlled by the choice of $S$ in the inner product Eq.~\eqref{eq_inner} that relates to the SNR in Eq.~\eqref{eq_SNR}.

\section{Application and results}\label{app_results}

We first report one selected analysis for a canonical analog model signal in Sec.~\ref{app_sch}. 
More details are then provided in Sec.~\ref{app_sch2}, and a noncanonical analog model signal is analyzed in Sec.~\ref{app_nsch}.

\subsection{Canonical analog model signal}\label{app_sch}

We first consider simulated data originating from perturbing the acoustic black hole effective metric from Eq.~\eqref{metric} to demonstrate and validate our framework. 
To balance the increasing complexity of the RZ metric but still allow for an efficient MCMC analysis, we consider a four-parameter model which shares the same RZ parameters as the injection given by
\begin{align}\label{theta4}
\bm{\theta^\mathrm{inj}} 
&= \left( r_0, a_0, b_0, \epsilon \right)
= \left( 1, 0, 0, 0\right)\,,
\end{align}
where throughout this framework we have fixed $a_1=b_1=0$, for both the injection and the sampling.

The efficiency of the RZ metric to describe a variety of non-Schwarzschild black holes has been demonstrated in the original work~\cite{Rezzolla:2014mua} and was further investigated in Ref.~\cite{Konoplya:2020hyk}. 
Therefore, we assume that keeping a similar number of free parameters is sufficient to model, in principle, various analog systems. 

\begin{figure}
\centering
\includegraphics[width=1.0\linewidth]{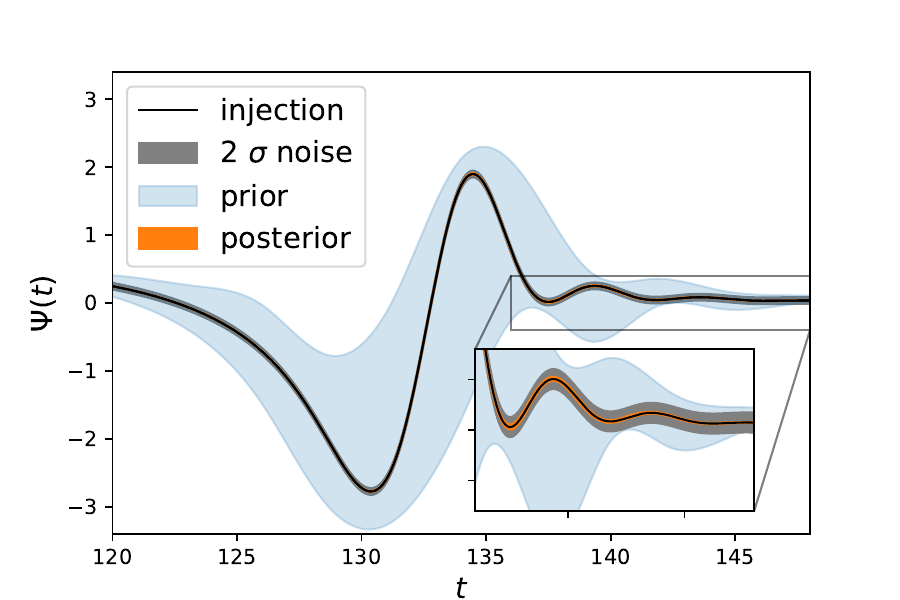}
\caption{Here, we show the injected signal (black solid) for $\bm\theta^\mathrm{inj}$. We also report the $95\,\%$ confidence levels obtained by sampling from the prior (blue) or the posterior (orange). 
Since we use a noiseless injection, we indicate the errors assumed in the likelihood by drawing $2 \sigma$ bands around the injection (gray). 
\label{signal2}
} 
\end{figure}

\begin{figure}
\centering
\includegraphics[width=1.0\linewidth]{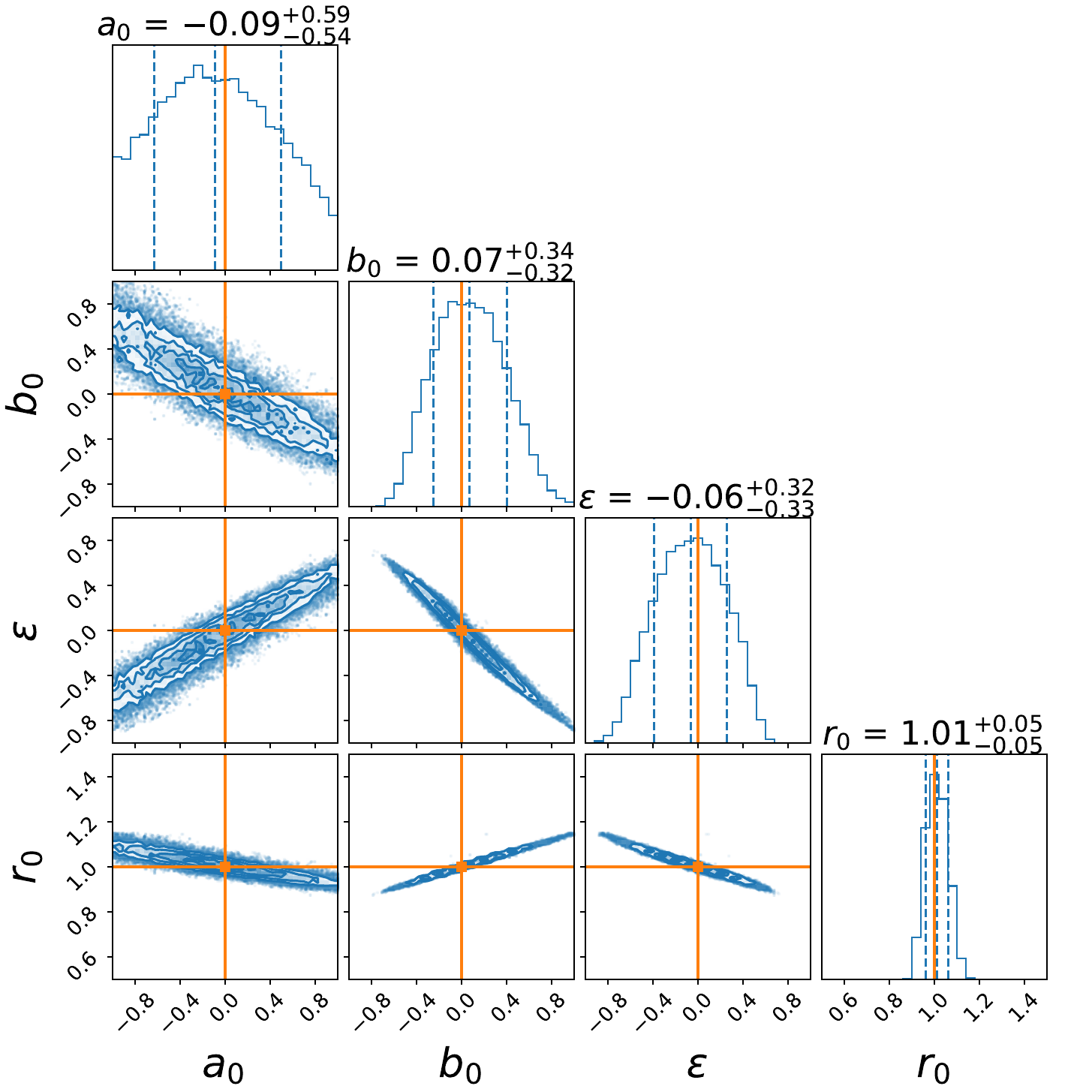}
\caption{Here we show the posterior distribution of four RZ parameters with injection $\bm\theta^\mathrm{inj}  = (r_0, a_0, b_0,\epsilon)=(1,0,0,0)$. 
Orange solid lines indicate the injected values, and blue dashed lines in the marginalized posteriors indicate quantiles of (0.16, 0.5, 0.84). 
\label{corner2}
}
\end{figure}

In Fig.~\ref{signal2}, we show the observed signal for $m=3$ produced by Gaussian initial data Eq.~\eqref{initialdata}. 
Note that $m=3$ is just one example, and that modern experiments could measure other $m$ as well~\cite{smaniotto2025blackholespectroscopygiantquantum}.
We consider the waveform strain $\bm{d}$ as measured by a distant observer at $r^{*}=80$ and set the SNR to be $\rho = 250$. 
Using the MCMC analysis with this signal, we can compute the posterior distribution $p(\bm{\theta}|\bm{d})$ of the RZ parameters reported in Fig.~\ref{corner2}. 
The colored regions in Fig.~\ref{signal2} and subsequent figures correspond to confidence levels showing the highest density interval of 95\,\% (or 68\,\%) when sampling from the prior or posterior distributions. 
All marginalized posteriors, shown on the diagonal panels, have their maximum very close to the injection $\bm\theta^\mathrm{inj}$, and all quantiles contain the injected parameters.

To get a better impression of how informative the posterior distribution is concerning the possible signals described by the prior, Fig.~\ref{signal2} also contains the $95\,\%$ confidence level obtained by sampling either from the prior and from the posterior. The graphics are generated by taking the $95\,\%$ highest density intervals from a large number of sampled $m(\bm \theta)$ curves. For the blue regions, $\bm \theta$ assumes random values from the prior distribution $p(\bm \theta )$, while for the orange areas $\bm \theta$ is sampled from the posterior distribution $p(\bm \theta | \bm d )$. The gray areas are obtained by adding and subtracting $2 \sigma$ from the injected signal $\bm d$.
These results demonstrate that the prior is much less informative than the posterior in describing the space of possible signals at late times but similar at early times. 
This is expected because early times mostly probe the metric's asymptotic properties, which are more constrained by construction.  

\begin{figure}
\centering
\includegraphics[width=1.0\linewidth]{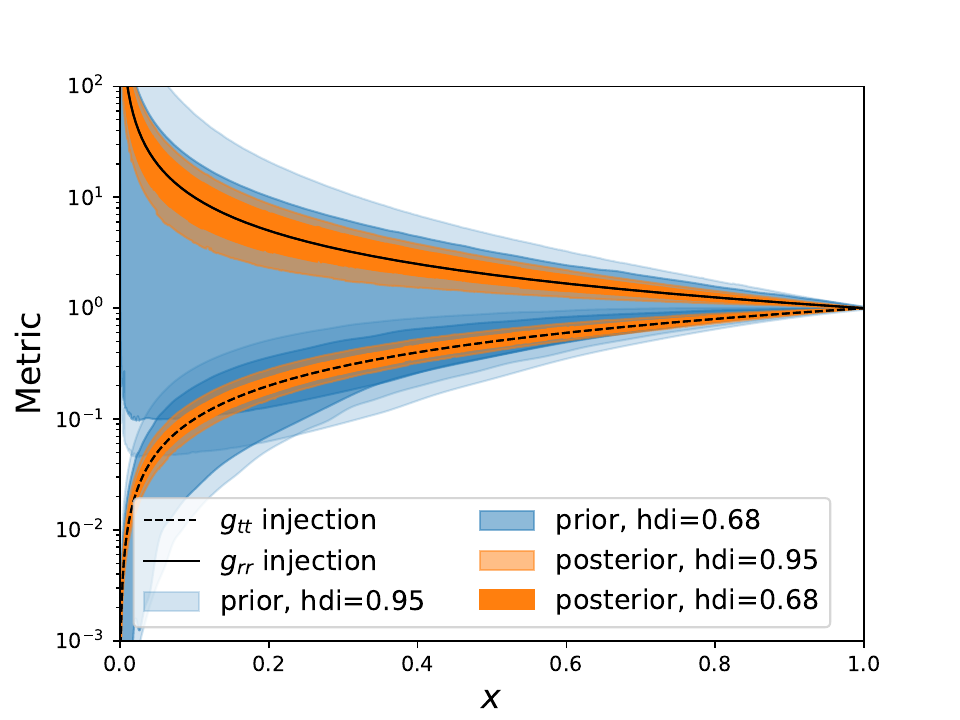}
\includegraphics[width=1.0\linewidth]{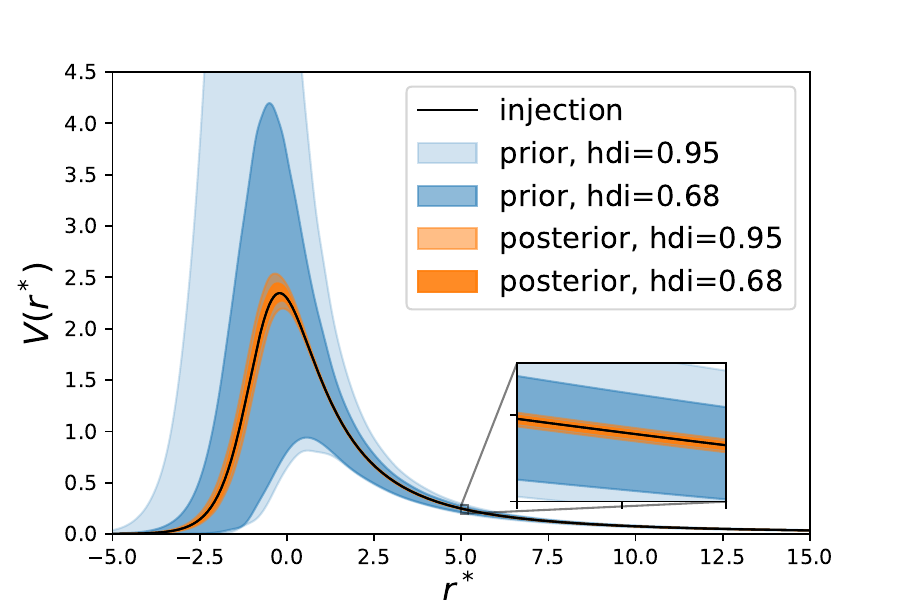}
\caption{Here, we show the prior and posterior distributions for the metric functions $g_{tt}$ and $g_{rr}$ in the top panel and the equivalent ones for the effective potential $V_m(r^*)$ in the bottom panel. 
We show the injection with the canonical analog model case and provide the $68\,\%$ and $95\,\%$ confidence levels obtained by sampling from the prior (blue) and the posterior (orange). 
\label{metric_potential2}
}
\end{figure}

Next, we show a similar comparison for the RZ metric functions $g_{tt}$ and $g_{rr}$ as well as the effective potential in Fig~\ref{metric_potential2}. 
Because $g_{rr}$ is described by one more parameter, it might not be surprising that its $95\,\%$ and $68\,\%$ confidence levels are larger than the ones for $g_{tt}$. The effective potential is well constrained for larger radii but becomes less constrained for radii smaller than the peak of the potential. 
This is expected because the RZ metric is more flexible there.

\subsection{Impact of SNR and observer location}\label{app_sch2}

Our previous example only considered one specific observer location and one SNR. In the following, we vary both aspects and repeat the MCMC analysis.   

\subsubsection{Impact of SNR}

To investigate how the SNR changes the posterior, we now include one smaller and one larger SNR $\rho=(100,500)$ to our analysis while fixing the observer location to $r^*_\mathrm{obs}=80$. 

\begin{figure}
\centering
\includegraphics[width=1.0\linewidth]{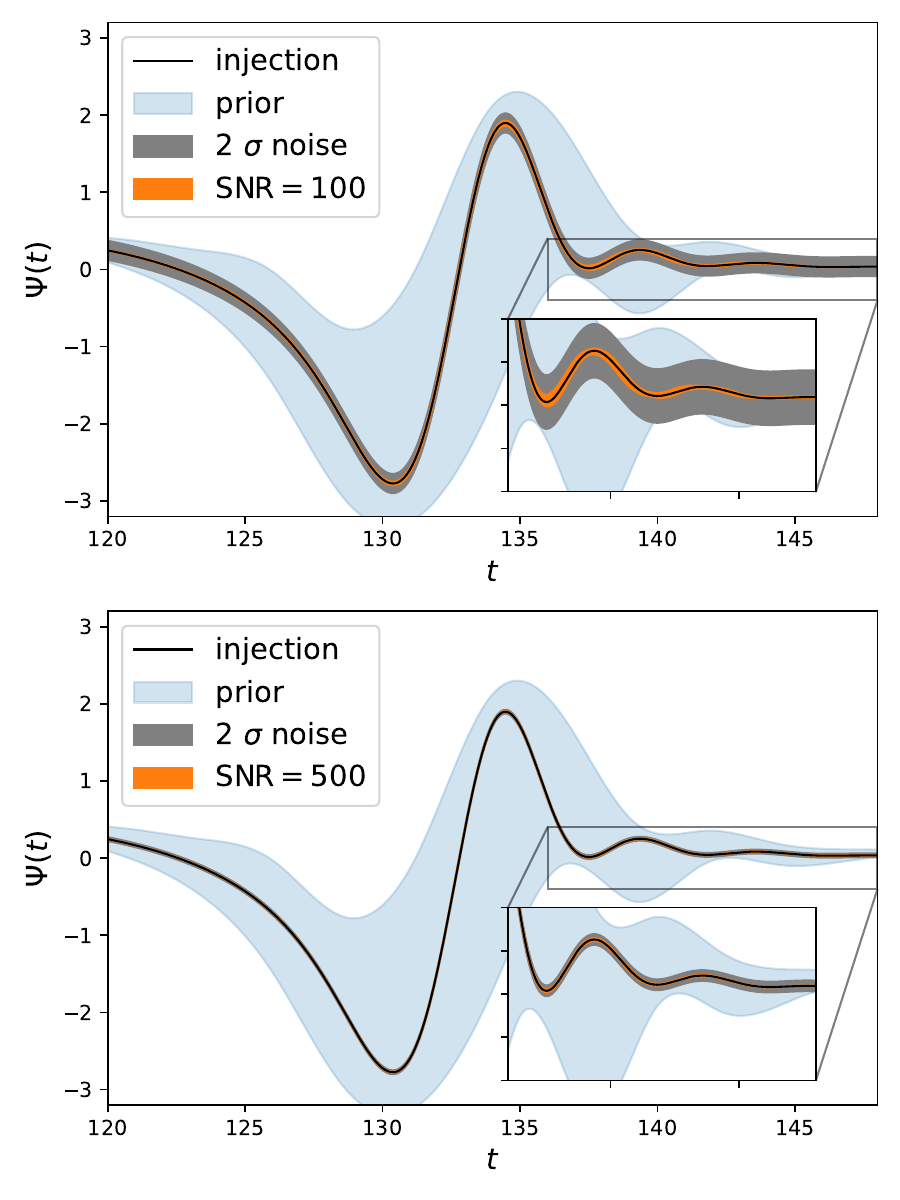}
\caption{Here, we show the injected signal (black solid) for $\bm\theta^\mathrm{inj}$ and the $95\,\%$ confidence levels obtained by sampling the priors (blue) and posteriors (orange) for SNRs of $\rho=100$ (upper panel) and $\rho=500$ (lower panel). 
The $2 \sigma$ noise levels are also indicated (gray). 
\label{signal3}
}
\end{figure}

In Fig.~\ref{signal3}, we report the injected signal with $2 \sigma$ noise levels, along with the highest density intervals of signals from sampling the prior $p(\bm{\theta})$ and posterior $p(\bm{\theta}|\bm{d})$ distributions of the RZ parameters. 
As expected, increasing the SNR results in more narrow posterior distributions as the data become more informative and less impacted by noise. 

\begin{figure}
\centering
\includegraphics[width=1.0\linewidth]{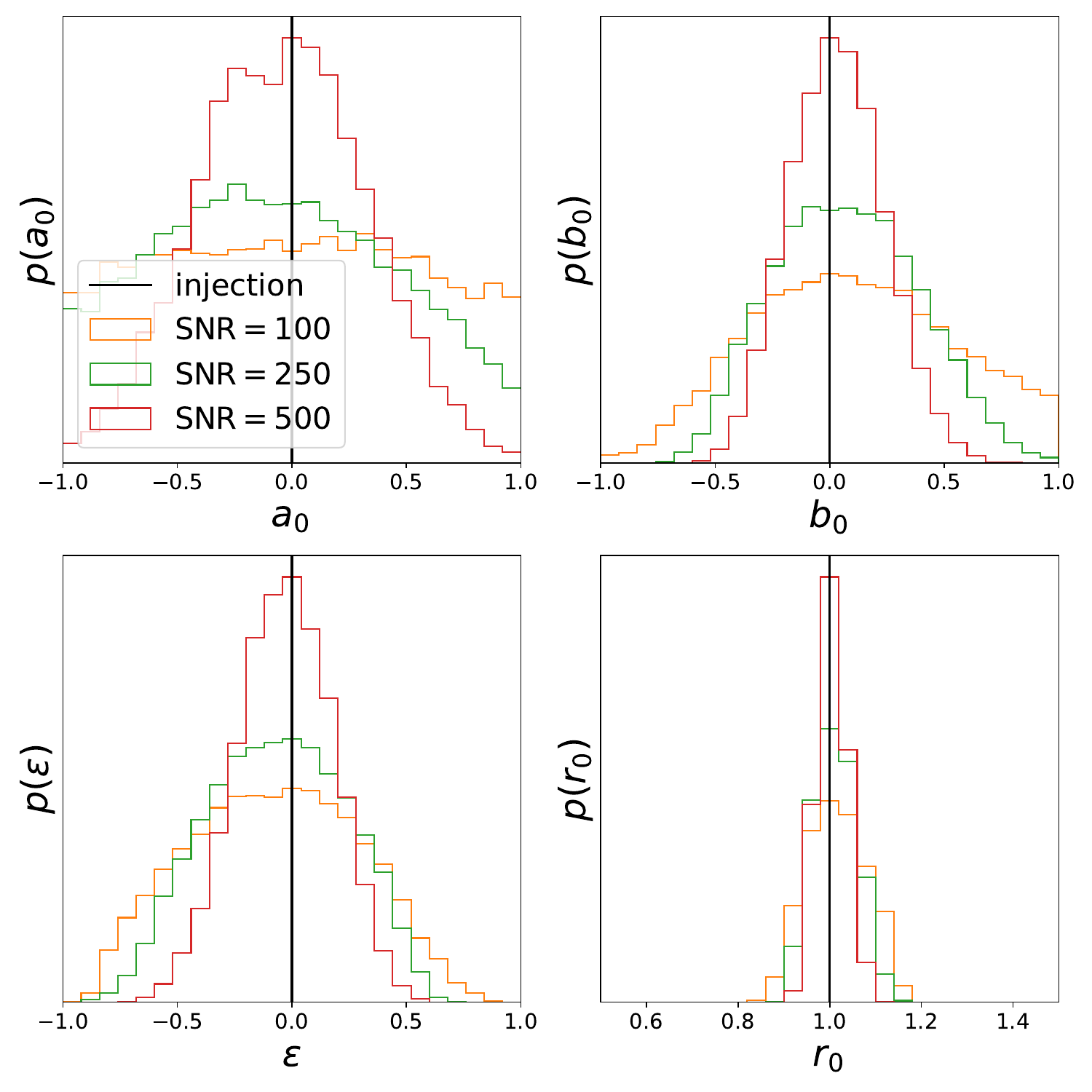}
\caption{Here, we show the marginalized posterior distributions of the four RZ parameters for an injected signal described by $\bm\theta^\mathrm{inj}$ (black solid lines) for three different SNRs. 
\label{corner3}
}
\end{figure}

The marginalized posteriors of the RZ parameters can be found in Fig.~\ref{corner3}, including the $\rho=250$ case from our previous application. 
While $\epsilon$ and $r_0$ are well localized within our prior range in all cases, $b_0$, and especially $a_0$, are overall less informed. 
Nevertheless, the injected values are well represented in all cases. 
Although it is nontrivial to quantitatively compare the obtained uncertainties of the different RZ parameters, note that $a_0$ is entering our realization of the RZ metric at the highest order. 
Thus, one would qualitatively expect it to be more sensitive to the potential closer to the horizon and, thus, harder to infer from the back-scattered wave. 

\begin{figure}
\centering
\includegraphics[width=1.0\linewidth]{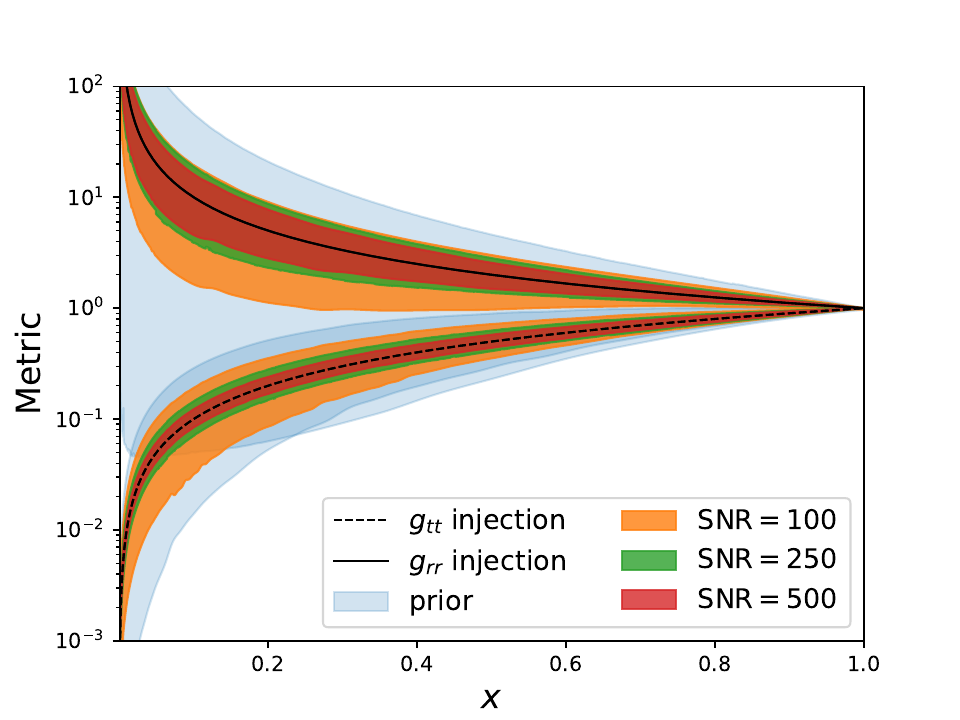}
\includegraphics[width=1.0\linewidth]{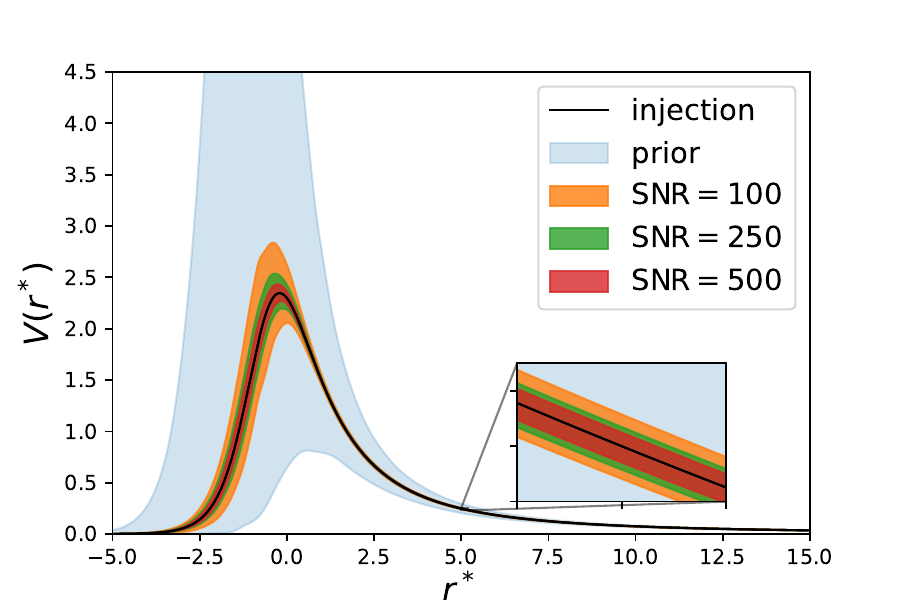}
\caption{Here, we show the posterior distributions for the metric functions $g_{tt}$ and $g_{rr}$ in the top panel and the equivalent ones for the effective potential $V_m(r^*)$ in the bottom panel. The posteriors correspond to three different SNRs.
}
\label{metric_potential3}
\end{figure}

Finally, to investigate the impact on the metric functions and effective potentials, we also show the corresponding highest density intervals in Fig.~\ref{metric_potential3}. 
All of them decrease with increasing SNR. 
Note that the effective potential depends in a non-trivial way on both metric functions. 
Therefore, the effective potential is more directly imprinted on the signal.

\subsubsection{Impact of observer distance}

Because the waveform can vary with the observer's location, one might expect that this could also affect the observer's sensitivity for inferring the physical parameters, at least as long one is not already in the far zone. 
In this case, only a phase shift is expected on the observed signal. 
We investigate this by varying the observer's distance from the source between $r^{*}_\mathrm{obs}=(80, 40, 10)$ while fixing the SNR to $\rho=100$, respectively. 

\begin{figure}
\centering
\includegraphics[width=1.0\linewidth]{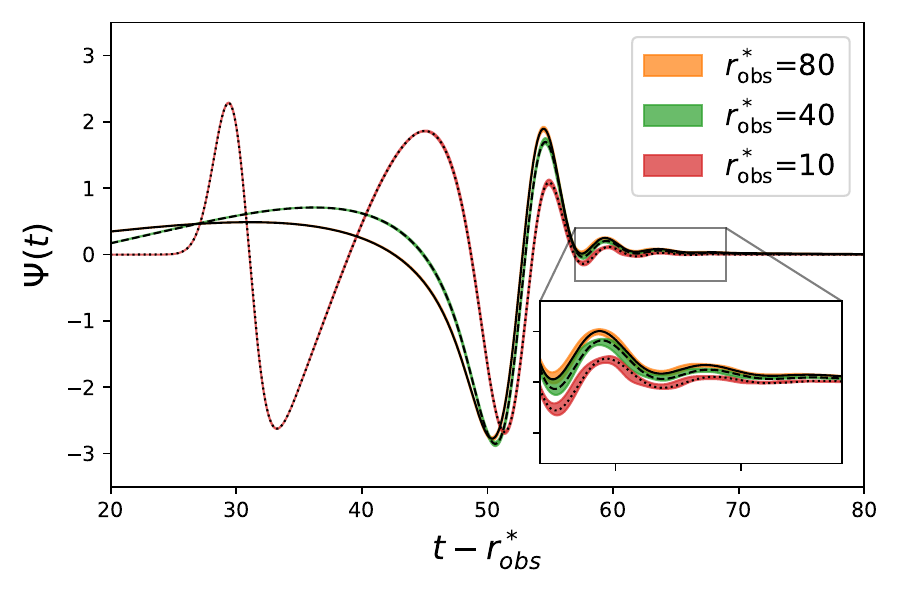}
\caption{Here, we show the injected signal (lines) for $\bm\theta^\mathrm{inj}$ with SNR=$100$ and the $95\,\%$ confidence levels obtained by sampling the posteriors from different observer locations. For a better visualization, we introduce a time shift of $t-r^*_{\text{obs}}$ for each waveform.
\label{signal4}
}
\end{figure}

\begin{figure}
\centering
\includegraphics[width=1.0\linewidth]{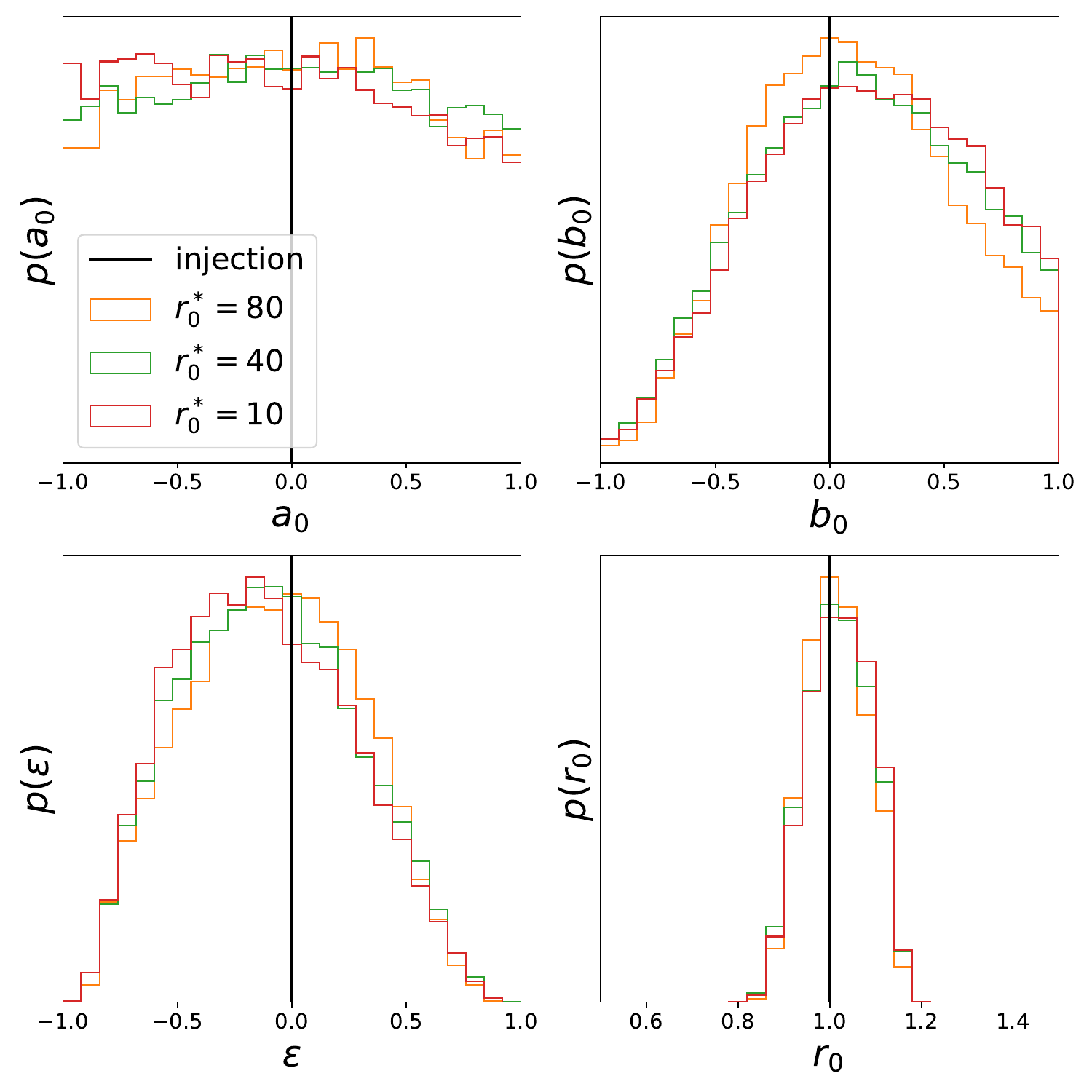}

\caption{Here, we show the marginalized posterior distributions of the four RZ parameters for an injected signal described by $\bm\theta^\mathrm{inj}$ (black solid lines) for three different observer positions. 
\label{corner4}
}
\end{figure}

In Fig.~\ref{signal4}, we provide the different waveforms as observed at different locations but shifted in time with respect to the observer location. 
Note that the waveforms at $r^{*}_\mathrm{obs}=40$ and $r^{*}_\mathrm{obs}=80$ are qualitatively similar, while the one at $r^{*}_\mathrm{obs}=10$ is quite different at early times, and only becomes similar at late times. 
The marginalized posterior distributions for the RZ parameters are reported in Fig.~\ref{corner4}. 
Despite the different waveform shapes at $r^{*}_\mathrm{obs}=10, 40$ and $80$, the results demonstrate that the inference is almost insensitive to it. 
While $a_0$ and $b_0$ are not well constrained compared to their priors, all observations are informative for determining $\varepsilon$ and $r_0$. 
Since the differences between the inferred metric functions and effective potentials are negligible, we do not explicitly show them. 

\subsection{Noncanonical analog model signal}\label{app_nsch}

To demonstrate that the inference framework is also capable of constraining possible deviations in analog gravity systems from their canonical description, we now consider simulated data originating from a nontrivial RZ injection given by
\begin{align}\label{theta4nGR}
\bm\theta^{ \mathrm{inj}\prime}  &= \left( r_0, a_0, b_0, \epsilon \right)
= \left(1, 0.3, 0.45, 0\right)\,.
\end{align}
As in Sec.~\ref{app_sch}, we fix the SNR to $\rho=250$ and the observer location at $r^{*}_\text{obs}=80$. 

\begin{figure}
\centering
\includegraphics[width=1.0\linewidth]{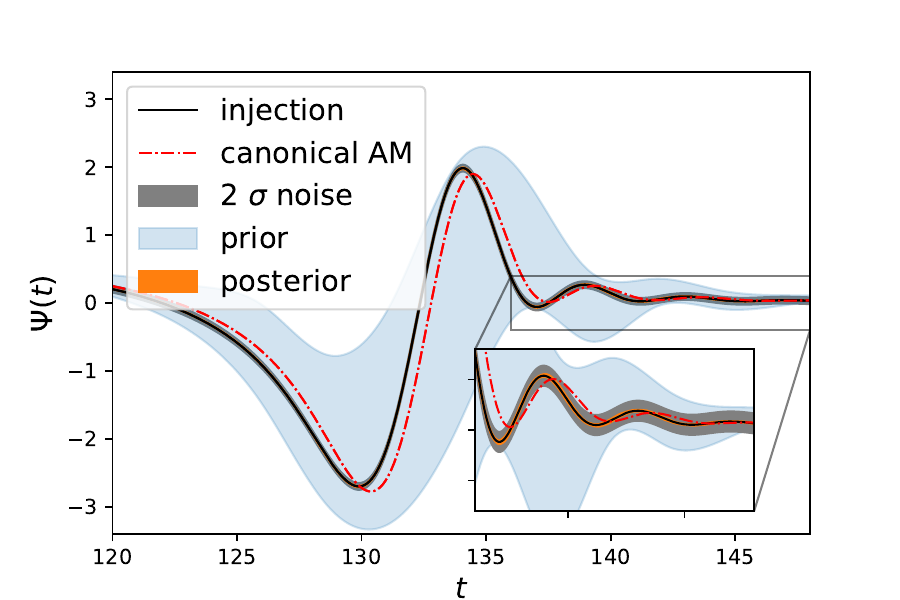}
\caption{Here, we show the injected signal (black solid) for $\bm\theta^{ \mathrm{inj}\prime}$ and the $95\,\%$ confidence levels obtained by sampling the posteriors for SNR=250 (orange) and the priors (blue). 
We also plot the injected signal for $\bm\theta^{ \mathrm{inj}}$ (red dashed) for comparison.
\label{signal5}
}
\end{figure}

\begin{figure}
\centering
\includegraphics[width=1.0\linewidth]{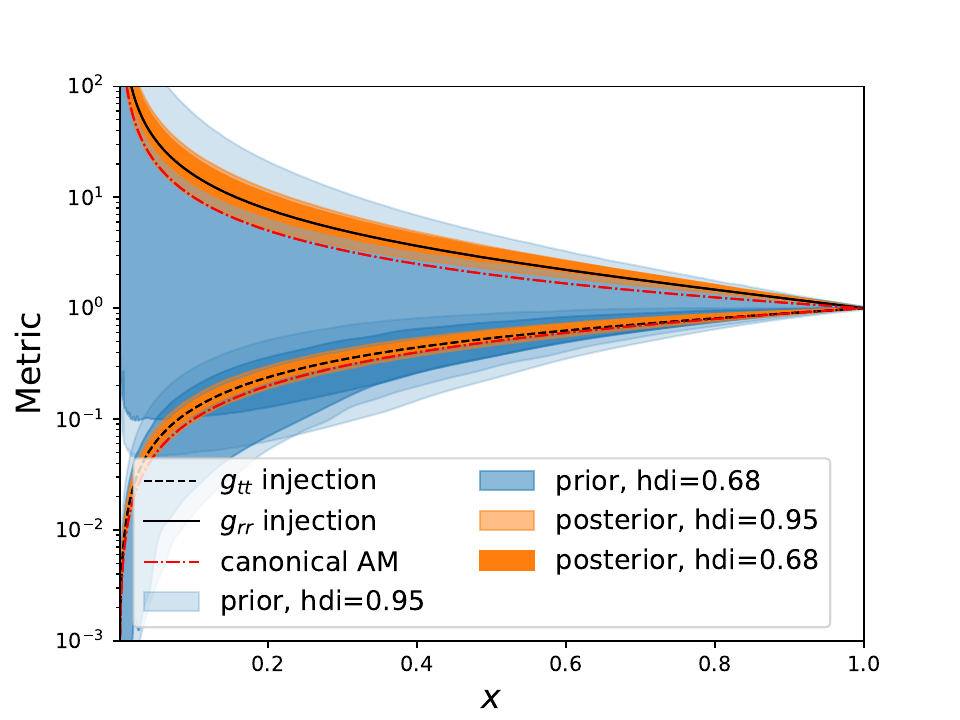}
\includegraphics[width=1.0\linewidth]{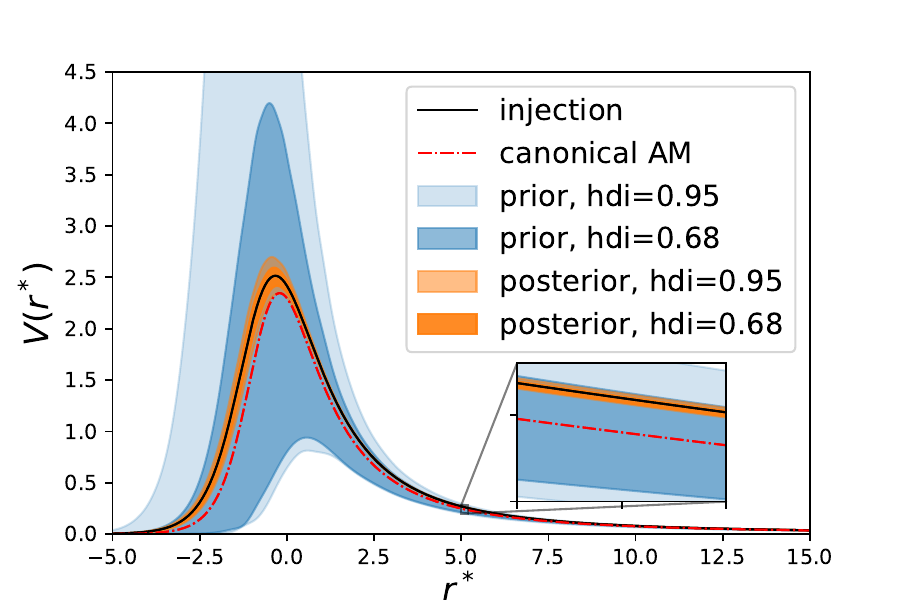}
\caption{Here, we show the posterior distributions for the metric functions $g_{tt}$ and $g_{rr}$ in the top panel and the equivalent ones for the effective potential $V_m(r^*)$ in the bottom panel. We also plot the injected metrics and potentials for $\bm\theta^{ \mathrm{inj}\prime}$ (black) and $\bm\theta^\mathrm{inj}$ (red dashed) for comparison.
\label{metric_potential5}
}
\end{figure}

In Fig.~\ref{signal5}, we show the injected signal $\bm{d}$ and the simulated samples for the prior and posterior distributions, with confidence level of $95\,\%$. 
For comparison, we also indicate the observed signal for the canonical scenario $\bm\theta^\mathrm{inj}$ studied earlier. 
Although the two injections look similar in shape, they differ in detail. 
Note that the canonical injection arrives slightly later than the non-canonical case. 
This indicates that the right side of the non-canonical effective potential barrier is overall shifted towards larger values of $r^*$, which can be confirmed in Fig.~\ref{metric_potential5}. 

\begin{figure}
\centering
\includegraphics[width=1.0\linewidth]{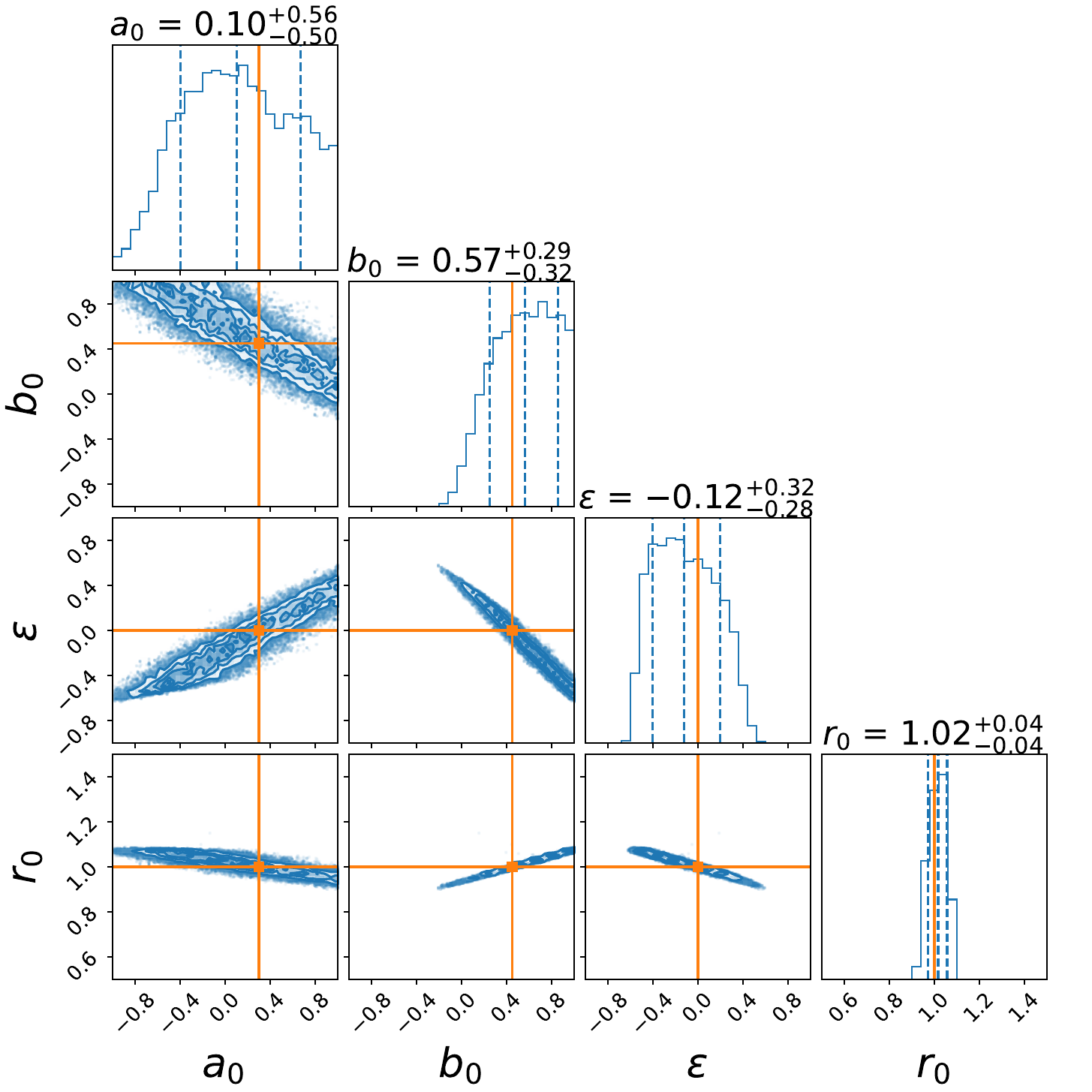}

\caption{Here, we show the posterior distribution of four RZ parameters with injection $\bm\theta^{ \mathrm{inj}\prime}  = (r_0, a_0, b_0,\epsilon)=(1,0.3,0.45,0)$. 
Orange solid lines indicate the injected values, and blue dashed lines in the marginalized posteriors indicate quantiles of (0.16,0.5, 0.84). 
\label{corner5}
}
\end{figure}

The posterior distribution of the RZ parameters is reported in Fig.~\ref{corner5}. 
Notice that the distributions are well centered around $\bm\theta^{ \mathrm{inj}\prime}$, as indicated by the quantiles, and show a more skewed behavior compared to the canonical case. 
We show the corresponding metric functions and effective potentials in Fig.~\ref{metric_potential5} for completeness. 
The canonical case lies mostly at the borders of the $95\,\%$ confidence level posteriors around the peak of the injected potential. Further out in the $r^*$ domain, the injection and the canonical potentials can be clearly distinguished from each other, as shown in the inset. 
This fact clearly demonstrates that the analysis is able to differentiate the noncanonical injection from our first example.

\section{Discussion}\label{discussion}

We discuss aspects of inference biases in Sec.~\ref{disc_biases}, compare our approach with related studies in Sec.~\ref{disc_otherworks}, and outline future extensions in Sec.~\ref{disc_future}.

\subsection{Remarks on inference biases}\label{disc_biases}

Performing fast sampling during the MCMC analysis requires evolving the wave propagation as efficiently as possible. 
There is a natural trade-off between refining the resolution of our numerical domain for accurate modeling and the time each step during sampling takes. 
However, the question of how accurately one must model data for parameter estimation depends crucially on the SNR. 
Simply speaking, if the noise in the signal is much larger than the numerical inaccuracies from the modeling, the width of the reconstructed posteriors will be broad and dominated by noise. 
Biases would be present if the SNR is high enough so that the analysis is sensitive to the details of the model. 
It would be visible by posteriors whose highest density intervals would not contain the injected parameters. 

Let us make two comments. 
First, in our work, we assume the same resolution for the injection and the modeling, which, by construction, avoids biases in a noiseless injection. 
Second, as a measure of how accurately our modeling waveforms capture the quasinormal mode ringing, we also extracted the fundamental quasinormal mode for the canonical analog model case and compared it to values known in the literature\cite{Matyjasek:2024uwo}. 
More specifically, we fitted a damped sinusoid to the exponentially decaying part of the injected signal. 
As a result, we obtained the real part of the fundamental mode with relative errors of 0.0005 and 0.0003 for its imaginary part.  

We note that using a higher resolution during sampling is, in principle, possible but would significantly increase the analysis time. 
To avoid this limitation, one could run a higher resolution simulation for a grid of the RZ parameters before sampling, store it, and interpolate between them later during sampling.

\subsection{Comparison to earlier works}\label{disc_otherworks}

The inverse problem related to analog gravity systems based on the Bohr-Sommerfeld rule and Gamow formula~\cite{lieb2015studies,MR985100,1980AmJPh..48..432L,2006AmJPh..74..638G} has been developed and studied in Refs.~\cite{Albuquerque:2023lzw,Albuquerque:2024xol,Albuquerque:2024cwl}. 
In these works, the input for reconstructing the perturbation potential and the boundary condition at the horizon was given by the transmission and reflection spectrum~\cite{Torres:2022bto}. 
In the case of partially reflecting boundary conditions, as suggested for models of exotic compact systems, the spectrum has a rich spectrum of resonances corresponding to long-lived trapped modes~\cite{doi:10.1098/rspa.1991.0104,Kokkotas:1994an}. 
In these works, it was necessary first to isolate these features and then feed them into the semianalytic equations for the reconstruction, as done for exotic compact objects in Refs.~\cite{Volkel:2017kfj,Volkel:2018hwb,Volkel:2019ahb}. 
One advantage of the semi-analytic approach is that it is parameter-free, i.e., one does not need to parametrize the effective potential or underlying metric, as we do in this work with the RZ metric. 
Both advantages and disadvantages are that the semiclassical method does not provide a unique solution. 
Instead, because the general problem is not uniquely solvable, it provides an approximation for the separation of turning points of the potential. 
The underlying assumption is the number of turning points in the effective potential.
One disadvantage is that the reconstruction process does not account for experimental noise, which could lead to instabilities in the reconstructed properties. 

Another study reconstructed the RZ metric using the quasinormal modes as input for a Bayesian analysis~\cite{Volkel:2020daa}. 
Since this study focuses on gravitational waves, it is impossible to apply the same time-domain analysis as carried out in this work. 
Instead, it is assumed that some quasinormal modes can be extracted and provided from an independent analysis. 
The quasinormal modes are then used as input for an analysis that directly computes the spectrum as a function of the RZ parameters. 
Realizations of the RZ metric that share partially similar quasinormal modes have been studied in Ref.~\cite{Volkel:2019muj}.

\subsection{Future extensions}\label{disc_future}

In the following, we discuss some of the many possible extensions of our work.

\subsubsection{Analyze full observer space}

In this work, we selected a few observers and analyzed their signals independently. 
However, one could also combine the data of multiple observers, and thus not only analyze the time-domain, but also resolve the full spatial domain. 
This would be qualitatively in the spirit of Ref.~\cite{Zhu:2023mzv}, which explores a so-called spacetime technique in the context of quasinormal mode extraction and explicitly using radial eigenfunctions, but not for noisy data. 
Since numerically computing the model $m(\bm{\theta})$ requires us to obtain the time series in the entire radial domain, it would be straightforward to include data from multiple observers in the likelihood. 
Since the main numerical cost consists of the wave propagation, not using it in the likelihood, it should be straightforward. 
However, realistic noise modeling may not be well approximated with Gaussian noise in the time domain, thus potentially making our likelihood evaluation more involved. 
Since this aspect will depend on the specific experiment, we leave it for future work.

\subsubsection{Combining different $m$}

In a real experiment, many different harmonics $m$ can be excited at the same time. 
Although it should be straightforward to extract the different harmonic contents $m$ to high accuracy, we have yet to explore how information of the RZ metric is imprinted across different $m$. 
Since the $m$ perturbation potentials are qualitatively similar and mainly increase height as a function of increasing $m$, we expect similar findings. 
However, we note that for large $m$, when approaching the eikonal limit, the potential is dominated by $g_{tt}$, and the impact of $g_{rr}$ becomes subdominant. 
Therefore, we expect measurements of the $b_i$ RZ parameters to become more challenging for large $m$. 
Regarding numerical cost, it would imply running the time evolution for each $m$ separately and would, therefore, slow down the MCMC analysis. 
We leave a quantitative analysis for future work.

\subsubsection{Varying the number of RZ parameters}

For simplicity, we chose a fixed number of RZ parameters for the injection and recovery. 
Since existing works demonstrate the capabilities of the RZ metric to provide a good approximation to non-Schwarzschild metrics by using only a few leading parameters~\cite{Rezzolla:2014mua,Konoplya:2020hyk}, we did not increase the number of RZ parameters. 
Including more parameters does not pose relevant problems to the performance of the time evolution; however, it will require one to include more samples for MCMC, thus making such analyses computationally more expensive. 
It would also require one to rule out nonphysical solutions of the RZ metric, e.g., via rejection during sampling. 
Future work could include varying the number of injected to reconstructed parameters or even applying it to exact analog black hole metrics that are not exactly represented by the RZ metric. 
Thus, the here presented approach can, in principle, be generalized to account for more complicated analog black hole metrics.

\subsubsection{Extension to rotating black holes}

The technically more challenging aspect will be to generalize our approach to rotating analog black holes. 
First, the background spacetime needs to be generalized, which could be readily done by using the KRZ metric~\cite{Konoplya:2016jvv}. 
However, the main problem one would face is that even the scalar field equation is, in general, not separable, which implies a more complicated set of coupled differential equations to be evolved. 
Although this by itself does not pose a fundamental problem, it would likely become the bottleneck when evolving it during sampling. 
One would thus be forced to develop an approximate representation for quick parameter estimation, i.e., similar to waveform models used for gravitational waves. 
It seems natural to follow the approach of waveform surrogates, which, in simple terms, span a well-chosen parameter space and perform full calculations on discrete points~\cite{Field:2013cfa,Purrer:2014fza}. 
Such models then utilize straightforward ways of interpolating between the parameters.

\subsubsection{Extension to analog gravity sampling techniques}

For acoustic black holes in Bose-Einstein condensates, it has been shown that the truncated Wigner method \cite{Sinatra_2002} can be used to study the existence of Hawking radiation  \cite{Carusotto:2008ep,Jacquet:2022vak}, and other analog effects \cite{10.21468/SciPostPhys.3.3.022,PhysRevLett.130.241501,PhysRevResearch.5.043282}. 
The method consists of studying the time evolution of perturbed samples, which may provide an interesting overlap between different sampling algorithms that could be explored in future work.

\section{Conclusions}\label{conclusions}

Analog gravity is an active field of research with fascinating connections between black hole physics and classical and quantum matter. 
In this work, we have presented a Bayesian analysis capable of reconstructing analog black hole metrics from time-evolution data corresponding to a draining vortex. 
As a universal ansatz for analog black holes of this type, we adjusted the RZ metric~\cite{Rezzolla:2014mua} to 2+1 dimensions. 
For the MCMC analysis, we utilized the \texttt{emcee} sampler~\cite{Foreman-Mackey:2012any}. 
To our knowledge, this is the first study of this kind.

Our setup is qualitatively related to the ringdown of binary black hole systems, an actively discussed topic in contemporary research. 
Among the open problems in a ringdown analysis with quasinormal modes is the need for precise knowledge of when the linearized regime where only the sum of quasinormal modes is a good approximation. 
Since the context of the analog gravity setup provides one with the initial data and, in principle, the possibility of choosing it, our analysis is free of such problems. 
Instead, we can analyze the entire signal, including the prompt response and the late-time power-law tails.

Future extensions of our work could open a range of possibilities, including furthering our understanding of the black hole ringdown. 
For example, one could compare the full ringdown analysis with a more traditional one based on the quasinormal mode spectrum. 
In this way, one could further quantify which part of the signal is sensitive to different regions of the effective potential and underlying metric. 
Other future directions are discussed in Sec.~\ref{disc_future}. 

On the analog gravity side, extending our framework to rotating black holes would be a technically challenging but conceptually straightforward objective. 
This would be closer to analog gravity experiments and more flexible for other applications. 
It would also be interesting to extend our framework to noisy initial data in the context of analog models for exotic compact objects~\cite{solidoro2024quasinormalmodessemiopensystems}. 
Another direction could be to look for possible applications concerning probing systems with a $1+1$ dimensional effective metric, such as those
generated in cold gases~\cite{keshet2024ringdownhawkingradiationconnectionreal}. 

\bigskip
\acknowledgments
The authors want to thank Kostas~D.~Kokkotas for useful discussions. 
The authors are also indebted to the anonymous referee for providing valuable comments that improved the presentation of this work.
S.~A. acknowledges funding from Conselho Nacional de Desenvolvimento Cient\'ifico e Tecnol\'ogico (CNPQ)-Brazil and Coordena\c{c}\~ao de Aperfei\c{c}oamento de Pessoal de N\'ivel Superior (CAPES)-Brazil. 
S.~H.~V. acknowledges funding from the Deutsche Forschungsgemeinschaft (DFG): Project No. 386119226. 

\bibliography{literature}

\begin{thebibliography}{105}%
\makeatletter
\providecommand \@ifxundefined [1]{%
 \@ifx{#1\undefined}
}%
\providecommand \@ifnum [1]{%
 \ifnum #1\expandafter \@firstoftwo
 \else \expandafter \@secondoftwo
 \fi
}%
\providecommand \@ifx [1]{%
 \ifx #1\expandafter \@firstoftwo
 \else \expandafter \@secondoftwo
 \fi
}%
\providecommand \natexlab [1]{#1}%
\providecommand \enquote  [1]{``#1''}%
\providecommand \bibnamefont  [1]{#1}%
\providecommand \bibfnamefont [1]{#1}%
\providecommand \citenamefont [1]{#1}%
\providecommand \href@noop [0]{\@secondoftwo}%
\providecommand \href [0]{\begingroup \@sanitize@url \@href}%
\providecommand \@href[1]{\@@startlink{#1}\@@href}%
\providecommand \@@href[1]{\endgroup#1\@@endlink}%
\providecommand \@sanitize@url [0]{\catcode `\\12\catcode `\$12\catcode `\&12\catcode `\#12\catcode `\^12\catcode `\_12\catcode `\%12\relax}%
\providecommand \@@startlink[1]{}%
\providecommand \@@endlink[0]{}%
\providecommand \url  [0]{\begingroup\@sanitize@url \@url }%
\providecommand \@url [1]{\endgroup\@href {#1}{\urlprefix }}%
\providecommand \urlprefix  [0]{URL }%
\providecommand \Eprint [0]{\href }%
\providecommand \doibase [0]{https://doi.org/}%
\providecommand \selectlanguage [0]{\@gobble}%
\providecommand \bibinfo  [0]{\@secondoftwo}%
\providecommand \bibfield  [0]{\@secondoftwo}%
\providecommand \translation [1]{[#1]}%
\providecommand \BibitemOpen [0]{}%
\providecommand \bibitemStop [0]{}%
\providecommand \bibitemNoStop [0]{.\EOS\space}%
\providecommand \EOS [0]{\spacefactor3000\relax}%
\providecommand \BibitemShut  [1]{\csname bibitem#1\endcsname}%
\let\auto@bib@innerbib\@empty
%</preamble>
\bibitem [{\citenamefont {Michell}(1784)}]{Michell:1784xqa}%
  \BibitemOpen
  \bibfield  {author} {\bibinfo {author} {\bibfnamefont {J.}~\bibnamefont {Michell}},\ }\bibfield  {title} {\bibinfo {title} {{On the Means of Discovering the Distance, Magnitude, \&c. of the Fixed Stars, in Consequence of the Diminution of the Velocity of Their Light, in Case Such a Diminution Should be Found to Take Place in any of Them, and Such Other Data Should be Procured from Observations, as Would be Farther Necessary for That Purpose.}},\ }\href {https://doi.org/10.1098/rstl.1784.0008} {\bibfield  {journal} {\bibinfo  {journal} {Phil. Trans. Roy. Soc. Lond.}\ }\textbf {\bibinfo {volume} {74}},\ \bibinfo {pages} {35} (\bibinfo {year} {1784})}\BibitemShut {NoStop}%
\bibitem [{\citenamefont {{Laplace}}(1799)}]{1799AllGE...4....1L}%
  \BibitemOpen
  \bibfield  {author} {\bibinfo {author} {\bibfnamefont {P.~S.}\ \bibnamefont {{Laplace}}},\ }\bibfield  {title} {\bibinfo {title} {{Beweis des Satzes, dass die anziehende Kraft bey einem Weltk{\"o}rper so gro{\ss} seyn k{\"o}nne, dass das Licht davon nicht ausstr{\"o}men kann}},\ }\href@noop {} {\bibfield  {journal} {\bibinfo  {journal} {Allgemeine Geographische Ephemeriden}\ }\textbf {\bibinfo {volume} {4}},\ \bibinfo {pages} {1} (\bibinfo {year} {1799})}\BibitemShut {NoStop}%
\bibitem [{\citenamefont {Einstein}(1916)}]{Einstein:1916vd}%
  \BibitemOpen
  \bibfield  {author} {\bibinfo {author} {\bibfnamefont {A.}~\bibnamefont {Einstein}},\ }\bibfield  {title} {\bibinfo {title} {{The foundation of the general theory of relativity.}},\ }\href {https://doi.org/10.1002/andp.19163540702} {\bibfield  {journal} {\bibinfo  {journal} {Annalen Phys.}\ }\textbf {\bibinfo {volume} {49}},\ \bibinfo {pages} {769} (\bibinfo {year} {1916})}\BibitemShut {NoStop}%
\bibitem [{\citenamefont {Schwarzschild}(1916)}]{Schwarzschild:1916uq}%
  \BibitemOpen
  \bibfield  {author} {\bibinfo {author} {\bibfnamefont {K.}~\bibnamefont {Schwarzschild}},\ }\bibfield  {title} {\bibinfo {title} {{On the gravitational field of a mass point according to Einstein's theory}},\ }\href@noop {} {\bibfield  {journal} {\bibinfo  {journal} {Sitzungsber. Preuss. Akad. Wiss. Berlin (Math. Phys. )}\ }\textbf {\bibinfo {volume} {1916}},\ \bibinfo {pages} {189} (\bibinfo {year} {1916})},\ \Eprint {https://arxiv.org/abs/physics/9905030} {arXiv:physics/9905030} \BibitemShut {NoStop}%
\bibitem [{\citenamefont {Kerr}(1963)}]{Kerr:1963ud}%
  \BibitemOpen
  \bibfield  {author} {\bibinfo {author} {\bibfnamefont {R.~P.}\ \bibnamefont {Kerr}},\ }\bibfield  {title} {\bibinfo {title} {{Gravitational field of a spinning mass as an example of algebraically special metrics}},\ }\href {https://doi.org/10.1103/PhysRevLett.11.237} {\bibfield  {journal} {\bibinfo  {journal} {Phys. Rev. Lett.}\ }\textbf {\bibinfo {volume} {11}},\ \bibinfo {pages} {237} (\bibinfo {year} {1963})}\BibitemShut {NoStop}%
\bibitem [{\citenamefont {Newman}\ \emph {et~al.}(1965)\citenamefont {Newman}, \citenamefont {Couch}, \citenamefont {Chinnapared}, \citenamefont {Exton}, \citenamefont {Prakash},\ and\ \citenamefont {Torrence}}]{Newman:1965my}%
  \BibitemOpen
  \bibfield  {author} {\bibinfo {author} {\bibfnamefont {E.~T.}\ \bibnamefont {Newman}}, \bibinfo {author} {\bibfnamefont {R.}~\bibnamefont {Couch}}, \bibinfo {author} {\bibfnamefont {K.}~\bibnamefont {Chinnapared}}, \bibinfo {author} {\bibfnamefont {A.}~\bibnamefont {Exton}}, \bibinfo {author} {\bibfnamefont {A.}~\bibnamefont {Prakash}},\ and\ \bibinfo {author} {\bibfnamefont {R.}~\bibnamefont {Torrence}},\ }\bibfield  {title} {\bibinfo {title} {{Metric of a Rotating, Charged Mass}},\ }\href {https://doi.org/10.1063/1.1704351} {\bibfield  {journal} {\bibinfo  {journal} {J. Math. Phys.}\ }\textbf {\bibinfo {volume} {6}},\ \bibinfo {pages} {918} (\bibinfo {year} {1965})}\BibitemShut {NoStop}%
\bibitem [{\citenamefont {Abbott}\ \emph {et~al.}(2016)\citenamefont {Abbott} \emph {et~al.}}]{LIGOScientific:2016aoc}%
  \BibitemOpen
  \bibfield  {author} {\bibinfo {author} {\bibfnamefont {B.~P.}\ \bibnamefont {Abbott}} \emph {et~al.} (\bibinfo {collaboration} {LIGO Scientific, Virgo}),\ }\bibfield  {title} {\bibinfo {title} {{Observation of Gravitational Waves from a Binary Black Hole Merger}},\ }\href {https://doi.org/10.1103/PhysRevLett.116.061102} {\bibfield  {journal} {\bibinfo  {journal} {Phys. Rev. Lett.}\ }\textbf {\bibinfo {volume} {116}},\ \bibinfo {pages} {061102} (\bibinfo {year} {2016})},\ \Eprint {https://arxiv.org/abs/1602.03837} {arXiv:1602.03837 [gr-qc]} \BibitemShut {NoStop}%
\bibitem [{\citenamefont {Akiyama}\ \emph {et~al.}(2019)\citenamefont {Akiyama} \emph {et~al.}}]{EventHorizonTelescope:2019dse}%
  \BibitemOpen
  \bibfield  {author} {\bibinfo {author} {\bibfnamefont {K.}~\bibnamefont {Akiyama}} \emph {et~al.} (\bibinfo {collaboration} {Event Horizon Telescope}),\ }\bibfield  {title} {\bibinfo {title} {{First M87 Event Horizon Telescope Results. I. The Shadow of the Supermassive Black Hole}},\ }\href {https://doi.org/10.3847/2041-8213/ab0ec7} {\bibfield  {journal} {\bibinfo  {journal} {Astrophys. J. Lett.}\ }\textbf {\bibinfo {volume} {875}},\ \bibinfo {pages} {L1} (\bibinfo {year} {2019})},\ \Eprint {https://arxiv.org/abs/1906.11238} {arXiv:1906.11238 [astro-ph.GA]} \BibitemShut {NoStop}%
\bibitem [{\citenamefont {Akiyama}\ \emph {et~al.}(2022)\citenamefont {Akiyama} \emph {et~al.}}]{EventHorizonTelescope:2022wkp}%
  \BibitemOpen
  \bibfield  {author} {\bibinfo {author} {\bibfnamefont {K.}~\bibnamefont {Akiyama}} \emph {et~al.} (\bibinfo {collaboration} {Event Horizon Telescope}),\ }\bibfield  {title} {\bibinfo {title} {{First Sagittarius A* Event Horizon Telescope Results. I. The Shadow of the Supermassive Black Hole in the Center of the Milky Way}},\ }\href {https://doi.org/10.3847/2041-8213/ac6674} {\bibfield  {journal} {\bibinfo  {journal} {Astrophys. J. Lett.}\ }\textbf {\bibinfo {volume} {930}},\ \bibinfo {pages} {L12} (\bibinfo {year} {2022})},\ \Eprint {https://arxiv.org/abs/2311.08680} {arXiv:2311.08680 [astro-ph.HE]} \BibitemShut {NoStop}%
\bibitem [{\citenamefont {Abuter}\ \emph {et~al.}(2020)\citenamefont {Abuter} \emph {et~al.}}]{GRAVITY:2020gka}%
  \BibitemOpen
  \bibfield  {author} {\bibinfo {author} {\bibfnamefont {R.}~\bibnamefont {Abuter}} \emph {et~al.} (\bibinfo {collaboration} {GRAVITY}),\ }\bibfield  {title} {\bibinfo {title} {{Detection of the Schwarzschild precession in the orbit of the star S2 near the Galactic centre massive black hole}},\ }\href {https://doi.org/10.1051/0004-6361/202037813} {\bibfield  {journal} {\bibinfo  {journal} {Astron. Astrophys.}\ }\textbf {\bibinfo {volume} {636}},\ \bibinfo {pages} {L5} (\bibinfo {year} {2020})},\ \Eprint {https://arxiv.org/abs/2004.07187} {arXiv:2004.07187 [astro-ph.GA]} \BibitemShut {NoStop}%
\bibitem [{\citenamefont {Bekenstein}(1973)}]{Bekenstein:1973ur}%
  \BibitemOpen
  \bibfield  {author} {\bibinfo {author} {\bibfnamefont {J.~D.}\ \bibnamefont {Bekenstein}},\ }\bibfield  {title} {\bibinfo {title} {{Black holes and entropy}},\ }\href {https://doi.org/10.1103/PhysRevD.7.2333} {\bibfield  {journal} {\bibinfo  {journal} {Phys. Rev. D}\ }\textbf {\bibinfo {volume} {7}},\ \bibinfo {pages} {2333} (\bibinfo {year} {1973})}\BibitemShut {NoStop}%
\bibitem [{\citenamefont {Bardeen}\ \emph {et~al.}(1973)\citenamefont {Bardeen}, \citenamefont {Carter},\ and\ \citenamefont {Hawking}}]{Bardeen:1973gs}%
  \BibitemOpen
  \bibfield  {author} {\bibinfo {author} {\bibfnamefont {J.~M.}\ \bibnamefont {Bardeen}}, \bibinfo {author} {\bibfnamefont {B.}~\bibnamefont {Carter}},\ and\ \bibinfo {author} {\bibfnamefont {S.~W.}\ \bibnamefont {Hawking}},\ }\bibfield  {title} {\bibinfo {title} {{The Four laws of black hole mechanics}},\ }\href {https://doi.org/10.1007/BF01645742} {\bibfield  {journal} {\bibinfo  {journal} {Commun. Math. Phys.}\ }\textbf {\bibinfo {volume} {31}},\ \bibinfo {pages} {161} (\bibinfo {year} {1973})}\BibitemShut {NoStop}%
\bibitem [{\citenamefont {Hawking}(1975)}]{Hawking:1975vcx}%
  \BibitemOpen
  \bibfield  {author} {\bibinfo {author} {\bibfnamefont {S.~W.}\ \bibnamefont {Hawking}},\ }\bibfield  {title} {\bibinfo {title} {{Particle Creation by Black Holes}},\ }\href {https://doi.org/10.1007/BF02345020} {\bibfield  {journal} {\bibinfo  {journal} {Commun. Math. Phys.}\ }\textbf {\bibinfo {volume} {43}},\ \bibinfo {pages} {199} (\bibinfo {year} {1975})},\ \bibinfo {note} {[Erratum: Commun.Math.Phys. 46, 206 (1976)]}\BibitemShut {NoStop}%
\bibitem [{\citenamefont {Barcelo}\ \emph {et~al.}(2005)\citenamefont {Barcelo}, \citenamefont {Liberati},\ and\ \citenamefont {Visser}}]{Barcelo:2005fc}%
  \BibitemOpen
  \bibfield  {author} {\bibinfo {author} {\bibfnamefont {C.}~\bibnamefont {Barcelo}}, \bibinfo {author} {\bibfnamefont {S.}~\bibnamefont {Liberati}},\ and\ \bibinfo {author} {\bibfnamefont {M.}~\bibnamefont {Visser}},\ }\bibfield  {title} {\bibinfo {title} {{Analogue gravity}},\ }\href {https://doi.org/10.12942/lrr-2005-12} {\bibfield  {journal} {\bibinfo  {journal} {Living Rev. Rel.}\ }\textbf {\bibinfo {volume} {8}},\ \bibinfo {pages} {12} (\bibinfo {year} {2005})},\ \Eprint {https://arxiv.org/abs/gr-qc/0505065} {arXiv:gr-qc/0505065} \BibitemShut {NoStop}%
\bibitem [{\citenamefont {Unruh}(1981)}]{Unruh:1980cg}%
  \BibitemOpen
  \bibfield  {author} {\bibinfo {author} {\bibfnamefont {W.~G.}\ \bibnamefont {Unruh}},\ }\bibfield  {title} {\bibinfo {title} {{Experimental black hole evaporation}},\ }\href {https://doi.org/10.1103/PhysRevLett.46.1351} {\bibfield  {journal} {\bibinfo  {journal} {Phys. Rev. Lett.}\ }\textbf {\bibinfo {volume} {46}},\ \bibinfo {pages} {1351} (\bibinfo {year} {1981})}\BibitemShut {NoStop}%
\bibitem [{\citenamefont {Visser}(1993)}]{visser1993acousticpropagationfluidsunexpected}%
  \BibitemOpen
  \bibfield  {author} {\bibinfo {author} {\bibfnamefont {M.}~\bibnamefont {Visser}},\ }\href {https://arxiv.org/abs/gr-qc/9311028} {\bibinfo {title} {Acoustic propagation in fluids: an unexpected example of lorentzian geometry}} (\bibinfo {year} {1993}),\ \Eprint {https://arxiv.org/abs/gr-qc/9311028} {arXiv:gr-qc/9311028 [gr-qc]} \BibitemShut {NoStop}%
\bibitem [{\citenamefont {Garay}\ \emph {et~al.}(2000)\citenamefont {Garay}, \citenamefont {Anglin}, \citenamefont {Cirac},\ and\ \citenamefont {Zoller}}]{bec1}%
  \BibitemOpen
  \bibfield  {author} {\bibinfo {author} {\bibfnamefont {L.~J.}\ \bibnamefont {Garay}}, \bibinfo {author} {\bibfnamefont {J.~R.}\ \bibnamefont {Anglin}}, \bibinfo {author} {\bibfnamefont {J.~I.}\ \bibnamefont {Cirac}},\ and\ \bibinfo {author} {\bibfnamefont {P.}~\bibnamefont {Zoller}},\ }\bibfield  {title} {\bibinfo {title} {{Sonic Analog of Gravitational Black Holes in Bose-Einstein Condensates}},\ }\href {https://doi.org/10.1103/PhysRevLett.85.4643} {\bibfield  {journal} {\bibinfo  {journal} {Phys. Rev. Lett.}\ }\textbf {\bibinfo {volume} {85}},\ \bibinfo {pages} {4643} (\bibinfo {year} {2000})}\BibitemShut {NoStop}%
\bibitem [{\citenamefont {Garay}\ \emph {et~al.}(2001)\citenamefont {Garay}, \citenamefont {Anglin}, \citenamefont {Cirac},\ and\ \citenamefont {Zoller}}]{bec2}%
  \BibitemOpen
  \bibfield  {author} {\bibinfo {author} {\bibfnamefont {L.~J.}\ \bibnamefont {Garay}}, \bibinfo {author} {\bibfnamefont {J.~R.}\ \bibnamefont {Anglin}}, \bibinfo {author} {\bibfnamefont {J.~I.}\ \bibnamefont {Cirac}},\ and\ \bibinfo {author} {\bibfnamefont {P.}~\bibnamefont {Zoller}},\ }\bibfield  {title} {\bibinfo {title} {{Sonic black holes in dilute Bose-Einstein condensates}},\ }\href {https://doi.org/10.1103/PhysRevA.63.023611} {\bibfield  {journal} {\bibinfo  {journal} {Phys. Rev. A}\ }\textbf {\bibinfo {volume} {63}},\ \bibinfo {pages} {023611} (\bibinfo {year} {2001})}\BibitemShut {NoStop}%
\bibitem [{\citenamefont {{Novello}}\ \emph {et~al.}(2002)\citenamefont {{Novello}}, \citenamefont {{Visser}},\ and\ \citenamefont {{Volovik}}}]{Novello2002}%
  \BibitemOpen
  \bibfield  {author} {\bibinfo {author} {\bibfnamefont {M.}~\bibnamefont {{Novello}}}, \bibinfo {author} {\bibfnamefont {M.}~\bibnamefont {{Visser}}},\ and\ \bibinfo {author} {\bibfnamefont {E.}~\bibnamefont {{Volovik}}, \bibfnamefont {Grigory}},\ }\href@noop {} {\emph {\bibinfo {title} {Artificial black holes}}}\ (\bibinfo  {publisher} {World Scientific},\ \bibinfo {address} {Hackensack, NJ},\ \bibinfo {year} {2002})\BibitemShut {NoStop}%
\bibitem [{\citenamefont {{Volovik}}(2003)}]{Volovik2003}%
  \BibitemOpen
  \bibfield  {author} {\bibinfo {author} {\bibfnamefont {G.~E.}\ \bibnamefont {{Volovik}}},\ }\href@noop {} {\emph {\bibinfo {title} {The Universe in a helium droplet}}}\ (\bibinfo  {publisher} {Oxford University Press},\ \bibinfo {address} {Oxford, UK},\ \bibinfo {year} {2003})\BibitemShut {NoStop}%
\bibitem [{\citenamefont {Nakano}\ \emph {et~al.}(2005)\citenamefont {Nakano}, \citenamefont {Kurita}, \citenamefont {Ogawa},\ and\ \citenamefont {Yoo}}]{Nakano:2004ha}%
  \BibitemOpen
  \bibfield  {author} {\bibinfo {author} {\bibfnamefont {H.}~\bibnamefont {Nakano}}, \bibinfo {author} {\bibfnamefont {Y.}~\bibnamefont {Kurita}}, \bibinfo {author} {\bibfnamefont {K.}~\bibnamefont {Ogawa}},\ and\ \bibinfo {author} {\bibfnamefont {C.-M.}\ \bibnamefont {Yoo}},\ }\bibfield  {title} {\bibinfo {title} {{Quasinormal ringing for acoustic black holes at low temperature}},\ }\href {https://doi.org/10.1103/PhysRevD.71.084006} {\bibfield  {journal} {\bibinfo  {journal} {Phys. Rev. D}\ }\textbf {\bibinfo {volume} {71}},\ \bibinfo {pages} {084006} (\bibinfo {year} {2005})},\ \Eprint {https://arxiv.org/abs/gr-qc/0411041} {arXiv:gr-qc/0411041} \BibitemShut {NoStop}%
\bibitem [{\citenamefont {{Unruh}}\ and\ \citenamefont {{Schützhold}}(2007)}]{Unruh2007}%
  \BibitemOpen
  \bibfield  {author} {\bibinfo {author} {\bibfnamefont {W.~G.}\ \bibnamefont {{Unruh}}}\ and\ \bibinfo {author} {\bibfnamefont {R.}~\bibnamefont {{Schützhold}}},\ }\href@noop {} {\emph {\bibinfo {title} {Quantum Analogues: From Phase Transitions to Black Holes and Cosmology}}}\ (\bibinfo  {publisher} {Springer},\ \bibinfo {address} {Berlin},\ \bibinfo {year} {2007})\BibitemShut {NoStop}%
\bibitem [{\citenamefont {Vieira}\ and\ \citenamefont {Kokkotas}(2021)}]{Vieira:2021xqw}%
  \BibitemOpen
  \bibfield  {author} {\bibinfo {author} {\bibfnamefont {H.~S.}\ \bibnamefont {Vieira}}\ and\ \bibinfo {author} {\bibfnamefont {K.~D.}\ \bibnamefont {Kokkotas}},\ }\bibfield  {title} {\bibinfo {title} {{Quasibound states of Schwarzschild acoustic black holes}},\ }\href {https://doi.org/10.1103/PhysRevD.104.024035} {\bibfield  {journal} {\bibinfo  {journal} {Phys. Rev. D}\ }\textbf {\bibinfo {volume} {104}},\ \bibinfo {pages} {024035} (\bibinfo {year} {2021})},\ \Eprint {https://arxiv.org/abs/2104.03938} {arXiv:2104.03938 [gr-qc]} \BibitemShut {NoStop}%
\bibitem [{\citenamefont {Steinhauer}(2014)}]{Steinhauer:2014dra}%
  \BibitemOpen
  \bibfield  {author} {\bibinfo {author} {\bibfnamefont {J.}~\bibnamefont {Steinhauer}},\ }\bibfield  {title} {\bibinfo {title} {{Observation of self-amplifying Hawking radiation in an analog black hole laser}},\ }\href {https://doi.org/10.1038/NPHYS3104} {\bibfield  {journal} {\bibinfo  {journal} {Nature Phys.}\ }\textbf {\bibinfo {volume} {10}},\ \bibinfo {pages} {864} (\bibinfo {year} {2014})},\ \Eprint {https://arxiv.org/abs/1409.6550} {arXiv:1409.6550 [cond-mat.quant-gas]} \BibitemShut {NoStop}%
\bibitem [{\citenamefont {Cardoso}\ \emph {et~al.}(2016)\citenamefont {Cardoso}, \citenamefont {Coutant}, \citenamefont {Richartz},\ and\ \citenamefont {Weinfurtner}}]{PhysRevLett.117.271101}%
  \BibitemOpen
  \bibfield  {author} {\bibinfo {author} {\bibfnamefont {V.}~\bibnamefont {Cardoso}}, \bibinfo {author} {\bibfnamefont {A.}~\bibnamefont {Coutant}}, \bibinfo {author} {\bibfnamefont {M.}~\bibnamefont {Richartz}},\ and\ \bibinfo {author} {\bibfnamefont {S.}~\bibnamefont {Weinfurtner}},\ }\bibfield  {title} {\bibinfo {title} {Detecting rotational superradiance in fluid laboratories},\ }\href {https://doi.org/10.1103/PhysRevLett.117.271101} {\bibfield  {journal} {\bibinfo  {journal} {Phys. Rev. Lett.}\ }\textbf {\bibinfo {volume} {117}},\ \bibinfo {pages} {271101} (\bibinfo {year} {2016})}\BibitemShut {NoStop}%
\bibitem [{\citenamefont {Barcel\'o}(2019)}]{Barcelo:2018ynq}%
  \BibitemOpen
  \bibfield  {author} {\bibinfo {author} {\bibfnamefont {C.}~\bibnamefont {Barcel\'o}},\ }\bibfield  {title} {\bibinfo {title} {{Analogue black-hole horizons}},\ }\href {https://doi.org/10.1038/s41567-018-0367-6} {\bibfield  {journal} {\bibinfo  {journal} {Nature Phys.}\ }\textbf {\bibinfo {volume} {15}},\ \bibinfo {pages} {210} (\bibinfo {year} {2019})}\BibitemShut {NoStop}%
\bibitem [{\citenamefont {{Vieira}}\ \emph {et~al.}(2022)\citenamefont {{Vieira}}, \citenamefont {{Destounis}},\ and\ \citenamefont {{Kokkotas}}}]{2022PhRvD.105d5015V}%
  \BibitemOpen
  \bibfield  {author} {\bibinfo {author} {\bibfnamefont {H.~S.}\ \bibnamefont {{Vieira}}}, \bibinfo {author} {\bibfnamefont {K.}~\bibnamefont {{Destounis}}},\ and\ \bibinfo {author} {\bibfnamefont {K.~D.}\ \bibnamefont {{Kokkotas}}},\ }\bibfield  {title} {\bibinfo {title} {{Slowly-rotating curved acoustic black holes: Quasinormal modes, Hawking-Unruh radiation, and quasibound states}},\ }\href {https://doi.org/10.1103/PhysRevD.105.045015} {\bibfield  {journal} {\bibinfo  {journal} {\prd}\ }\textbf {\bibinfo {volume} {105}},\ \bibinfo {eid} {045015} (\bibinfo {year} {2022})},\ \Eprint {https://arxiv.org/abs/2112.08711} {arXiv:2112.08711 [gr-qc]} \BibitemShut {NoStop}%
\bibitem [{\citenamefont {Vieira}\ \emph {et~al.}(2023)\citenamefont {Vieira}, \citenamefont {Destounis},\ and\ \citenamefont {Kokkotas}}]{kostashoracio}%
  \BibitemOpen
  \bibfield  {author} {\bibinfo {author} {\bibfnamefont {H.~S.}\ \bibnamefont {Vieira}}, \bibinfo {author} {\bibfnamefont {K.}~\bibnamefont {Destounis}},\ and\ \bibinfo {author} {\bibfnamefont {K.~D.}\ \bibnamefont {Kokkotas}},\ }\bibfield  {title} {\bibinfo {title} {{Analog Schwarzschild black holes of Bose-Einstein condensates in a cavity: Quasinormal modes and quasibound states}},\ }\href {https://doi.org/10.1103/PhysRevD.107.104038} {\bibfield  {journal} {\bibinfo  {journal} {Phys. Rev. D}\ }\textbf {\bibinfo {volume} {107}},\ \bibinfo {pages} {104038} (\bibinfo {year} {2023})}\BibitemShut {NoStop}%
\bibitem [{\citenamefont {Solidoro}\ \emph {et~al.}(2024)\citenamefont {Solidoro}, \citenamefont {Patrick}, \citenamefont {Gregory},\ and\ \citenamefont {Weinfurtner}}]{solidoro2024quasinormalmodessemiopensystems}%
  \BibitemOpen
  \bibfield  {author} {\bibinfo {author} {\bibfnamefont {L.}~\bibnamefont {Solidoro}}, \bibinfo {author} {\bibfnamefont {S.}~\bibnamefont {Patrick}}, \bibinfo {author} {\bibfnamefont {R.}~\bibnamefont {Gregory}},\ and\ \bibinfo {author} {\bibfnamefont {S.}~\bibnamefont {Weinfurtner}},\ }\href {https://arxiv.org/abs/2406.11013} {\bibinfo {title} {Quasinormal modes in semi-open systems}} (\bibinfo {year} {2024}),\ \Eprint {https://arxiv.org/abs/2406.11013} {arXiv:2406.11013 [gr-qc]} \BibitemShut {NoStop}%
\bibitem [{\citenamefont {Del~Porro}\ \emph {et~al.}(2025)\citenamefont {Del~Porro}, \citenamefont {Liberati},\ and\ \citenamefont {Schneider}}]{DelPorro:2024tuw}%
  \BibitemOpen
  \bibfield  {author} {\bibinfo {author} {\bibfnamefont {F.}~\bibnamefont {Del~Porro}}, \bibinfo {author} {\bibfnamefont {S.}~\bibnamefont {Liberati}},\ and\ \bibinfo {author} {\bibfnamefont {M.}~\bibnamefont {Schneider}},\ }\bibfield  {title} {\bibinfo {title} {{Tunneling method for Hawking quanta in analogue gravity}},\ }\href {https://doi.org/10.5802/crphys.239} {\bibfield  {journal} {\bibinfo  {journal} {Comptes Rendus Physique}\ }\textbf {\bibinfo {volume} {25}},\ \bibinfo {pages} {1} (\bibinfo {year} {2025})},\ \Eprint {https://arxiv.org/abs/2406.14603} {arXiv:2406.14603 [gr-qc]} \BibitemShut {NoStop}%
\bibitem [{\citenamefont {Keshet}\ \emph {et~al.}(2024)\citenamefont {Keshet}, \citenamefont {Shemesh},\ and\ \citenamefont {Steinhauer}}]{keshet2024ringdownhawkingradiationconnectionreal}%
  \BibitemOpen
  \bibfield  {author} {\bibinfo {author} {\bibfnamefont {E.}~\bibnamefont {Keshet}}, \bibinfo {author} {\bibfnamefont {I.}~\bibnamefont {Shemesh}},\ and\ \bibinfo {author} {\bibfnamefont {J.}~\bibnamefont {Steinhauer}},\ }\href {https://arxiv.org/abs/2407.00448} {\bibinfo {title} {The ringdown-hawking radiation connection in real and analogue black holes}} (\bibinfo {year} {2024}),\ \Eprint {https://arxiv.org/abs/2407.00448} {arXiv:2407.00448 [gr-qc]} \BibitemShut {NoStop}%
\bibitem [{\citenamefont {Liao}\ \emph {et~al.}(2019)\citenamefont {Liao}, \citenamefont {van~der Wurff}, \citenamefont {van Oosten},\ and\ \citenamefont {Stoof}}]{Liao:2018avv}%
  \BibitemOpen
  \bibfield  {author} {\bibinfo {author} {\bibfnamefont {L.}~\bibnamefont {Liao}}, \bibinfo {author} {\bibfnamefont {E.~C.~I.}\ \bibnamefont {van~der Wurff}}, \bibinfo {author} {\bibfnamefont {D.}~\bibnamefont {van Oosten}},\ and\ \bibinfo {author} {\bibfnamefont {H.~T.~C.}\ \bibnamefont {Stoof}},\ }\bibfield  {title} {\bibinfo {title} {{Proposal for an analog Schwarzschild black hole in condensates of light}},\ }\href {https://doi.org/10.1103/PhysRevA.99.023850} {\bibfield  {journal} {\bibinfo  {journal} {Phys. Rev. A}\ }\textbf {\bibinfo {volume} {99}},\ \bibinfo {pages} {023850} (\bibinfo {year} {2019})},\ \Eprint {https://arxiv.org/abs/1806.00023} {arXiv:1806.00023 [cond-mat.quant-gas]} \BibitemShut {NoStop}%
\bibitem [{\citenamefont {Moreno-Ruiz}\ and\ \citenamefont {Bermudez}(2022)}]{Moreno-Ruiz:2021qrf}%
  \BibitemOpen
  \bibfield  {author} {\bibinfo {author} {\bibfnamefont {A.}~\bibnamefont {Moreno-Ruiz}}\ and\ \bibinfo {author} {\bibfnamefont {D.}~\bibnamefont {Bermudez}},\ }\bibfield  {title} {\bibinfo {title} {{Optical analogue of the Schwarzschild\textendash{}Planck metric}},\ }\href {https://doi.org/10.1088/1361-6382/ac7506} {\bibfield  {journal} {\bibinfo  {journal} {Class. Quant. Grav.}\ }\textbf {\bibinfo {volume} {39}},\ \bibinfo {pages} {145001} (\bibinfo {year} {2022})},\ \Eprint {https://arxiv.org/abs/2112.00194} {arXiv:2112.00194 [gr-qc]} \BibitemShut {NoStop}%
\bibitem [{\citenamefont {Haller}\ \emph {et~al.}(2023)\citenamefont {Haller}, \citenamefont {Hegde}, \citenamefont {Xu}, \citenamefont {De~Beule}, \citenamefont {Schmidt},\ and\ \citenamefont {Meng}}]{Haller:2022pfj}%
  \BibitemOpen
  \bibfield  {author} {\bibinfo {author} {\bibfnamefont {A.}~\bibnamefont {Haller}}, \bibinfo {author} {\bibfnamefont {S.}~\bibnamefont {Hegde}}, \bibinfo {author} {\bibfnamefont {C.}~\bibnamefont {Xu}}, \bibinfo {author} {\bibfnamefont {C.}~\bibnamefont {De~Beule}}, \bibinfo {author} {\bibfnamefont {T.~L.}\ \bibnamefont {Schmidt}},\ and\ \bibinfo {author} {\bibfnamefont {T.}~\bibnamefont {Meng}},\ }\bibfield  {title} {\bibinfo {title} {{Black hole mirages: Electron lensing and Berry curvature effects in inhomogeneously tilted Weyl semimetals}},\ }\href {https://doi.org/10.21468/SciPostPhys.14.5.119} {\bibfield  {journal} {\bibinfo  {journal} {SciPost Phys.}\ }\textbf {\bibinfo {volume} {14}},\ \bibinfo {pages} {119} (\bibinfo {year} {2023})},\ \Eprint {https://arxiv.org/abs/2210.16254} {arXiv:2210.16254 [cond-mat.mes-hall]} \BibitemShut {NoStop}%
\bibitem [{\citenamefont {Smaniotto}\ \emph {et~al.}(2025)\citenamefont {Smaniotto}, \citenamefont {Solidoro}, \citenamefont {Švančara}, \citenamefont {Patrick}, \citenamefont {Richartz}, \citenamefont {Barenghi}, \citenamefont {Gregory},\ and\ \citenamefont {Weinfurtner}}]{smaniotto2025blackholespectroscopygiantquantum}%
  \BibitemOpen
  \bibfield  {author} {\bibinfo {author} {\bibfnamefont {P.}~\bibnamefont {Smaniotto}}, \bibinfo {author} {\bibfnamefont {L.}~\bibnamefont {Solidoro}}, \bibinfo {author} {\bibfnamefont {P.}~\bibnamefont {Švančara}}, \bibinfo {author} {\bibfnamefont {S.}~\bibnamefont {Patrick}}, \bibinfo {author} {\bibfnamefont {M.}~\bibnamefont {Richartz}}, \bibinfo {author} {\bibfnamefont {C.~F.}\ \bibnamefont {Barenghi}}, \bibinfo {author} {\bibfnamefont {R.}~\bibnamefont {Gregory}},\ and\ \bibinfo {author} {\bibfnamefont {S.}~\bibnamefont {Weinfurtner}},\ }\href {https://arxiv.org/abs/2502.11209} {\bibinfo {title} {Black-hole spectroscopy from a giant quantum vortex}} (\bibinfo {year} {2025}),\ \Eprint {https://arxiv.org/abs/2502.11209} {arXiv:2502.11209 [gr-qc]} \BibitemShut {NoStop}%
\bibitem [{\citenamefont {\ifmmode \check{C}\else \v{C}\fi{}love\ifmmode~\check{c}\else \v{c}\fi{}ko}\ \emph {et~al.}(2019)\citenamefont {\ifmmode \check{C}\else \v{C}\fi{}love\ifmmode~\check{c}\else \v{c}\fi{}ko}, \citenamefont {Ga\ifmmode~\check{z}\else \v{z}\fi{}o}, \citenamefont {Kupka},\ and\ \citenamefont {Skyba}}]{PhysRevLett.123.161302}%
  \BibitemOpen
  \bibfield  {author} {\bibinfo {author} {\bibfnamefont {M.}~\bibnamefont {\ifmmode \check{C}\else \v{C}\fi{}love\ifmmode~\check{c}\else \v{c}\fi{}ko}}, \bibinfo {author} {\bibfnamefont {E.}~\bibnamefont {Ga\ifmmode~\check{z}\else \v{z}\fi{}o}}, \bibinfo {author} {\bibfnamefont {M.}~\bibnamefont {Kupka}},\ and\ \bibinfo {author} {\bibfnamefont {P.}~\bibnamefont {Skyba}},\ }\bibfield  {title} {\bibinfo {title} {Magnonic analog of black- and white-hole horizons in superfluid $^{3}\mathrm{He}\text{\ensuremath{-}}b$},\ }\href {https://doi.org/10.1103/PhysRevLett.123.161302} {\bibfield  {journal} {\bibinfo  {journal} {Phys. Rev. Lett.}\ }\textbf {\bibinfo {volume} {123}},\ \bibinfo {pages} {161302} (\bibinfo {year} {2019})}\BibitemShut {NoStop}%
\bibitem [{\citenamefont {Steinhauer}(2016)}]{Steinhauer:2015saa}%
  \BibitemOpen
  \bibfield  {author} {\bibinfo {author} {\bibfnamefont {J.}~\bibnamefont {Steinhauer}},\ }\bibfield  {title} {\bibinfo {title} {{Observation of quantum Hawking radiation and its entanglement in an analogue black hole}},\ }\href {https://doi.org/10.1038/nphys3863} {\bibfield  {journal} {\bibinfo  {journal} {Nature Phys.}\ }\textbf {\bibinfo {volume} {12}},\ \bibinfo {pages} {959} (\bibinfo {year} {2016})},\ \Eprint {https://arxiv.org/abs/1510.00621} {arXiv:1510.00621 [gr-qc]} \BibitemShut {NoStop}%
\bibitem [{\citenamefont {\v{C}love\v{c}ko}\ \emph {et~al.}(2019)\citenamefont {\v{C}love\v{c}ko}, \citenamefont {Ga\v{z}o}, \citenamefont {Kupka},\ and\ \citenamefont {Skyba}}]{Clovecko:2018qnj}%
  \BibitemOpen
  \bibfield  {author} {\bibinfo {author} {\bibfnamefont {M.}~\bibnamefont {\v{C}love\v{c}ko}}, \bibinfo {author} {\bibfnamefont {E.}~\bibnamefont {Ga\v{z}o}}, \bibinfo {author} {\bibfnamefont {M.}~\bibnamefont {Kupka}},\ and\ \bibinfo {author} {\bibfnamefont {P.}~\bibnamefont {Skyba}},\ }\bibfield  {title} {\bibinfo {title} {{Magnonic Analog of Black- and White-Hole Horizons in Superfluid $^3$He-B}},\ }\href {https://doi.org/10.1103/PhysRevLett.123.161302} {\bibfield  {journal} {\bibinfo  {journal} {Phys. Rev. Lett.}\ }\textbf {\bibinfo {volume} {123}},\ \bibinfo {pages} {161302} (\bibinfo {year} {2019})},\ \Eprint {https://arxiv.org/abs/1810.09890} {arXiv:1810.09890 [cond-mat.other]} \BibitemShut {NoStop}%
\bibitem [{\citenamefont {Braunstein}\ \emph {et~al.}(2023)\citenamefont {Braunstein}, \citenamefont {Faizal}, \citenamefont {Krauss}, \citenamefont {Marino},\ and\ \citenamefont {Shah}}]{Braunstein:2023jpo}%
  \BibitemOpen
  \bibfield  {author} {\bibinfo {author} {\bibfnamefont {S.~L.}\ \bibnamefont {Braunstein}}, \bibinfo {author} {\bibfnamefont {M.}~\bibnamefont {Faizal}}, \bibinfo {author} {\bibfnamefont {L.~M.}\ \bibnamefont {Krauss}}, \bibinfo {author} {\bibfnamefont {F.}~\bibnamefont {Marino}},\ and\ \bibinfo {author} {\bibfnamefont {N.~A.}\ \bibnamefont {Shah}},\ }\bibfield  {title} {\bibinfo {title} {{Analogue simulations of quantum gravity with fluids}},\ }\href {https://doi.org/10.1038/s42254-023-00630-y} {\bibfield  {journal} {\bibinfo  {journal} {Nature Rev. Phys.}\ }\textbf {\bibinfo {volume} {5}},\ \bibinfo {pages} {612} (\bibinfo {year} {2023})},\ \Eprint {https://arxiv.org/abs/2402.16136} {arXiv:2402.16136 [gr-qc]} \BibitemShut {NoStop}%
\bibitem [{\citenamefont {\v{S}van\v{c}ara}\ \emph {et~al.}(2024)\citenamefont {\v{S}van\v{c}ara}, \citenamefont {Smaniotto}, \citenamefont {Solidoro}, \citenamefont {MacDonald}, \citenamefont {Patrick}, \citenamefont {Gregory}, \citenamefont {Barenghi},\ and\ \citenamefont {Weinfurtner}}]{Svancara:2023yrf}%
  \BibitemOpen
  \bibfield  {author} {\bibinfo {author} {\bibfnamefont {P.}~\bibnamefont {\v{S}van\v{c}ara}}, \bibinfo {author} {\bibfnamefont {P.}~\bibnamefont {Smaniotto}}, \bibinfo {author} {\bibfnamefont {L.}~\bibnamefont {Solidoro}}, \bibinfo {author} {\bibfnamefont {J.~F.}\ \bibnamefont {MacDonald}}, \bibinfo {author} {\bibfnamefont {S.}~\bibnamefont {Patrick}}, \bibinfo {author} {\bibfnamefont {R.}~\bibnamefont {Gregory}}, \bibinfo {author} {\bibfnamefont {C.~F.}\ \bibnamefont {Barenghi}},\ and\ \bibinfo {author} {\bibfnamefont {S.}~\bibnamefont {Weinfurtner}},\ }\bibfield  {title} {\bibinfo {title} {{Rotating curved spacetime signatures from a giant quantum vortex}},\ }\href {https://doi.org/10.1038/s41586-024-07176-8} {\bibfield  {journal} {\bibinfo  {journal} {Nature}\ }\textbf {\bibinfo {volume} {628}},\ \bibinfo {pages} {66} (\bibinfo {year} {2024})},\ \Eprint {https://arxiv.org/abs/2308.10773} {arXiv:2308.10773 [gr-qc]} \BibitemShut {NoStop}%
\bibitem [{\citenamefont {Fischer}\ and\ \citenamefont {Sch\"utzhold}(2004)}]{extra1}%
  \BibitemOpen
  \bibfield  {author} {\bibinfo {author} {\bibfnamefont {U.~R.}\ \bibnamefont {Fischer}}\ and\ \bibinfo {author} {\bibfnamefont {R.}~\bibnamefont {Sch\"utzhold}},\ }\bibfield  {title} {\bibinfo {title} {Quantum simulation of cosmic inflation in two-component bose-einstein condensates},\ }\href {https://doi.org/10.1103/PhysRevA.70.063615} {\bibfield  {journal} {\bibinfo  {journal} {Phys. Rev. A}\ }\textbf {\bibinfo {volume} {70}},\ \bibinfo {pages} {063615} (\bibinfo {year} {2004})}\BibitemShut {NoStop}%
\bibitem [{\citenamefont {Fedichev}\ and\ \citenamefont {Fischer}(2004)}]{extra2}%
  \BibitemOpen
  \bibfield  {author} {\bibinfo {author} {\bibfnamefont {P.~O.}\ \bibnamefont {Fedichev}}\ and\ \bibinfo {author} {\bibfnamefont {U.~R.}\ \bibnamefont {Fischer}},\ }\bibfield  {title} {\bibinfo {title} {``cosmological'' quasiparticle production in harmonically trapped superfluid gases},\ }\href {https://doi.org/10.1103/PhysRevA.69.033602} {\bibfield  {journal} {\bibinfo  {journal} {Phys. Rev. A}\ }\textbf {\bibinfo {volume} {69}},\ \bibinfo {pages} {033602} (\bibinfo {year} {2004})}\BibitemShut {NoStop}%
\bibitem [{\citenamefont {Ribeiro}\ \emph {et~al.}(2022)\citenamefont {Ribeiro}, \citenamefont {Baak},\ and\ \citenamefont {Fischer}}]{extra3}%
  \BibitemOpen
  \bibfield  {author} {\bibinfo {author} {\bibfnamefont {C.~C.~H.}\ \bibnamefont {Ribeiro}}, \bibinfo {author} {\bibfnamefont {S.-S.}\ \bibnamefont {Baak}},\ and\ \bibinfo {author} {\bibfnamefont {U.~R.}\ \bibnamefont {Fischer}},\ }\bibfield  {title} {\bibinfo {title} {Existence of steady-state black hole analogs in finite quasi-one-dimensional bose-einstein condensates},\ }\href {https://doi.org/10.1103/PhysRevD.105.124066} {\bibfield  {journal} {\bibinfo  {journal} {Phys. Rev. D}\ }\textbf {\bibinfo {volume} {105}},\ \bibinfo {pages} {124066} (\bibinfo {year} {2022})}\BibitemShut {NoStop}%
\bibitem [{\citenamefont {Holanda~Ribeiro}\ and\ \citenamefont {Fischer}(2023)}]{extra4}%
  \BibitemOpen
  \bibfield  {author} {\bibinfo {author} {\bibfnamefont {C.~C.}\ \bibnamefont {Holanda~Ribeiro}}\ and\ \bibinfo {author} {\bibfnamefont {U.~R.}\ \bibnamefont {Fischer}},\ }\bibfield  {title} {\bibinfo {title} {Impact of trans-planckian excitations on black-hole radiation in dipolar condensates},\ }\href {https://doi.org/10.1103/PhysRevD.107.L121502} {\bibfield  {journal} {\bibinfo  {journal} {Phys. Rev. D}\ }\textbf {\bibinfo {volume} {107}},\ \bibinfo {pages} {L121502} (\bibinfo {year} {2023})}\BibitemShut {NoStop}%
\bibitem [{\citenamefont {Barcel\'{o}}\ \emph {et~al.}(2003)\citenamefont {Barcel\'{o}}, \citenamefont {Liberati},\ and\ \citenamefont {Visser}}]{hawkingbec1}%
  \BibitemOpen
  \bibfield  {author} {\bibinfo {author} {\bibfnamefont {C.}~\bibnamefont {Barcel\'{o}}}, \bibinfo {author} {\bibfnamefont {S.}~\bibnamefont {Liberati}},\ and\ \bibinfo {author} {\bibfnamefont {M.}~\bibnamefont {Visser}},\ }\bibfield  {title} {\bibinfo {title} {{Towards the Observation of Hawking Radiation in Bose–Einstein Condensates}},\ }\href {https://doi.org/10.1142/S0217751X0301615X} {\bibfield  {journal} {\bibinfo  {journal} {International Journal of Modern Physics A}\ }\textbf {\bibinfo {volume} {18}},\ \bibinfo {pages} {3735} (\bibinfo {year} {2003})}\BibitemShut {NoStop}%
\bibitem [{\citenamefont {Barcel\'o}\ \emph {et~al.}(2003)\citenamefont {Barcel\'o}, \citenamefont {Liberati},\ and\ \citenamefont {Visser}}]{hawkingbec2}%
  \BibitemOpen
  \bibfield  {author} {\bibinfo {author} {\bibfnamefont {C.}~\bibnamefont {Barcel\'o}}, \bibinfo {author} {\bibfnamefont {S.}~\bibnamefont {Liberati}},\ and\ \bibinfo {author} {\bibfnamefont {M.}~\bibnamefont {Visser}},\ }\bibfield  {title} {\bibinfo {title} {{Probing semiclassical analog gravity in Bose-Einstein condensates with widely tunable interactions}},\ }\href {https://doi.org/10.1103/PhysRevA.68.053613} {\bibfield  {journal} {\bibinfo  {journal} {Phys. Rev. A}\ }\textbf {\bibinfo {volume} {68}},\ \bibinfo {pages} {053613} (\bibinfo {year} {2003})}\BibitemShut {NoStop}%
\bibitem [{\citenamefont {Fedichev}\ and\ \citenamefont {Fischer}(2003)}]{hawkingbec3}%
  \BibitemOpen
  \bibfield  {author} {\bibinfo {author} {\bibfnamefont {P.~O.}\ \bibnamefont {Fedichev}}\ and\ \bibinfo {author} {\bibfnamefont {U.~R.}\ \bibnamefont {Fischer}},\ }\bibfield  {title} {\bibinfo {title} {{Gibbons-Hawking Effect in the Sonic de Sitter Space-Time of an Expanding Bose-Einstein-Condensed Gas}},\ }\href {https://doi.org/10.1103/PhysRevLett.91.240407} {\bibfield  {journal} {\bibinfo  {journal} {Phys. Rev. Lett.}\ }\textbf {\bibinfo {volume} {91}},\ \bibinfo {pages} {240407} (\bibinfo {year} {2003})}\BibitemShut {NoStop}%
\bibitem [{\citenamefont {Rousseaux}\ \emph {et~al.}(2008)\citenamefont {Rousseaux}, \citenamefont {Mathis}, \citenamefont {Maissa}, \citenamefont {Philbin},\ and\ \citenamefont {Leonhardt}}]{Rousseaux:2007is}%
  \BibitemOpen
  \bibfield  {author} {\bibinfo {author} {\bibfnamefont {G.}~\bibnamefont {Rousseaux}}, \bibinfo {author} {\bibfnamefont {C.}~\bibnamefont {Mathis}}, \bibinfo {author} {\bibfnamefont {P.}~\bibnamefont {Maissa}}, \bibinfo {author} {\bibfnamefont {T.~G.}\ \bibnamefont {Philbin}},\ and\ \bibinfo {author} {\bibfnamefont {U.}~\bibnamefont {Leonhardt}},\ }\bibfield  {title} {\bibinfo {title} {{Observation of negative phase velocity waves in a water tank: A classical analogue to the Hawking effect?}},\ }\href {https://doi.org/10.1088/1367-2630/10/5/053015} {\bibfield  {journal} {\bibinfo  {journal} {New J. Phys.}\ }\textbf {\bibinfo {volume} {10}},\ \bibinfo {pages} {053015} (\bibinfo {year} {2008})},\ \Eprint {https://arxiv.org/abs/0711.4767} {arXiv:0711.4767 [gr-qc]} \BibitemShut {NoStop}%
\bibitem [{\citenamefont {Euv\'e}\ \emph {et~al.}(2016)\citenamefont {Euv\'e}, \citenamefont {Michel}, \citenamefont {Parentani}, \citenamefont {Philbin},\ and\ \citenamefont {Rousseaux}}]{PhysRevLett.117.121301}%
  \BibitemOpen
  \bibfield  {author} {\bibinfo {author} {\bibfnamefont {L.-P.}\ \bibnamefont {Euv\'e}}, \bibinfo {author} {\bibfnamefont {F.}~\bibnamefont {Michel}}, \bibinfo {author} {\bibfnamefont {R.}~\bibnamefont {Parentani}}, \bibinfo {author} {\bibfnamefont {T.~G.}\ \bibnamefont {Philbin}},\ and\ \bibinfo {author} {\bibfnamefont {G.}~\bibnamefont {Rousseaux}},\ }\bibfield  {title} {\bibinfo {title} {Observation of noise correlated by the hawking effect in a water tank},\ }\href {https://doi.org/10.1103/PhysRevLett.117.121301} {\bibfield  {journal} {\bibinfo  {journal} {Phys. Rev. Lett.}\ }\textbf {\bibinfo {volume} {117}},\ \bibinfo {pages} {121301} (\bibinfo {year} {2016})}\BibitemShut {NoStop}%
\bibitem [{\citenamefont {{Mu{\~n}oz de Nova}}\ \emph {et~al.}(2019)\citenamefont {{Mu{\~n}oz de Nova}}, \citenamefont {{Golubkov}}, \citenamefont {{Kolobov}},\ and\ \citenamefont {{Steinhauer}}}]{2019Natur.569..688M}%
  \BibitemOpen
  \bibfield  {author} {\bibinfo {author} {\bibfnamefont {J.~R.}\ \bibnamefont {{Mu{\~n}oz de Nova}}}, \bibinfo {author} {\bibfnamefont {K.}~\bibnamefont {{Golubkov}}}, \bibinfo {author} {\bibfnamefont {V.~I.}\ \bibnamefont {{Kolobov}}},\ and\ \bibinfo {author} {\bibfnamefont {J.}~\bibnamefont {{Steinhauer}}},\ }\bibfield  {title} {\bibinfo {title} {{Observation of thermal Hawking radiation and its temperature in an analogue black hole}},\ }\href {https://doi.org/10.1038/s41586-019-1241-0} {\bibfield  {journal} {\bibinfo  {journal} {\nat}\ }\textbf {\bibinfo {volume} {569}},\ \bibinfo {pages} {688} (\bibinfo {year} {2019})},\ \Eprint {https://arxiv.org/abs/1809.00913} {arXiv:1809.00913 [gr-qc]} \BibitemShut {NoStop}%
\bibitem [{\citenamefont {{Kolobov}}\ \emph {et~al.}(2021)\citenamefont {{Kolobov}}, \citenamefont {{Golubkov}}, \citenamefont {{Mu{\~n}oz de Nova}},\ and\ \citenamefont {{Steinhauer}}}]{2021NatPh..17..362K}%
  \BibitemOpen
  \bibfield  {author} {\bibinfo {author} {\bibfnamefont {V.~I.}\ \bibnamefont {{Kolobov}}}, \bibinfo {author} {\bibfnamefont {K.}~\bibnamefont {{Golubkov}}}, \bibinfo {author} {\bibfnamefont {J.~R.}\ \bibnamefont {{Mu{\~n}oz de Nova}}},\ and\ \bibinfo {author} {\bibfnamefont {J.}~\bibnamefont {{Steinhauer}}},\ }\bibfield  {title} {\bibinfo {title} {{Observation of stationary spontaneous Hawking radiation and the time evolution of an analogue black hole}},\ }\href {https://doi.org/10.1038/s41567-020-01076-0} {\bibfield  {journal} {\bibinfo  {journal} {Nature Physics}\ }\textbf {\bibinfo {volume} {17}},\ \bibinfo {pages} {362} (\bibinfo {year} {2021})},\ \Eprint {https://arxiv.org/abs/1910.09363} {arXiv:1910.09363 [gr-qc]} \BibitemShut {NoStop}%
\bibitem [{\citenamefont {{Torres}}\ \emph {et~al.}(2017)\citenamefont {{Torres}}, \citenamefont {{Patrick}}, \citenamefont {{Coutant}}, \citenamefont {{Richartz}}, \citenamefont {{Tedford}},\ and\ \citenamefont {{Weinfurtner}}}]{2017NatPh..13..833T}%
  \BibitemOpen
  \bibfield  {author} {\bibinfo {author} {\bibfnamefont {T.}~\bibnamefont {{Torres}}}, \bibinfo {author} {\bibfnamefont {S.}~\bibnamefont {{Patrick}}}, \bibinfo {author} {\bibfnamefont {A.}~\bibnamefont {{Coutant}}}, \bibinfo {author} {\bibfnamefont {M.}~\bibnamefont {{Richartz}}}, \bibinfo {author} {\bibfnamefont {E.~W.}\ \bibnamefont {{Tedford}}},\ and\ \bibinfo {author} {\bibfnamefont {S.}~\bibnamefont {{Weinfurtner}}},\ }\bibfield  {title} {\bibinfo {title} {{Rotational superradiant scattering in a vortex flow}},\ }\href {https://doi.org/10.1038/nphys4151} {\bibfield  {journal} {\bibinfo  {journal} {Nature Physics}\ }\textbf {\bibinfo {volume} {13}},\ \bibinfo {pages} {833} (\bibinfo {year} {2017})}\BibitemShut {NoStop}%
\bibitem [{\citenamefont {Torres}\ \emph {et~al.}(2020)\citenamefont {Torres}, \citenamefont {Patrick}, \citenamefont {Richartz},\ and\ \citenamefont {Weinfurtner}}]{Torres:2020tzs}%
  \BibitemOpen
  \bibfield  {author} {\bibinfo {author} {\bibfnamefont {T.}~\bibnamefont {Torres}}, \bibinfo {author} {\bibfnamefont {S.}~\bibnamefont {Patrick}}, \bibinfo {author} {\bibfnamefont {M.}~\bibnamefont {Richartz}},\ and\ \bibinfo {author} {\bibfnamefont {S.}~\bibnamefont {Weinfurtner}},\ }\bibfield  {title} {\bibinfo {title} {{Quasinormal Mode Oscillations in an Analogue Black Hole Experiment}},\ }\href {https://doi.org/10.1103/PhysRevLett.125.011301} {\bibfield  {journal} {\bibinfo  {journal} {Phys. Rev. Lett.}\ }\textbf {\bibinfo {volume} {125}},\ \bibinfo {pages} {011301} (\bibinfo {year} {2020})},\ \Eprint {https://arxiv.org/abs/1811.07858} {arXiv:1811.07858 [gr-qc]} \BibitemShut {NoStop}%
\bibitem [{\citenamefont {Rezzolla}\ and\ \citenamefont {Zhidenko}(2014)}]{Rezzolla:2014mua}%
  \BibitemOpen
  \bibfield  {author} {\bibinfo {author} {\bibfnamefont {L.}~\bibnamefont {Rezzolla}}\ and\ \bibinfo {author} {\bibfnamefont {A.}~\bibnamefont {Zhidenko}},\ }\bibfield  {title} {\bibinfo {title} {{New parametrization for spherically symmetric black holes in metric theories of gravity}},\ }\href {https://doi.org/10.1103/PhysRevD.90.084009} {\bibfield  {journal} {\bibinfo  {journal} {Phys. Rev. D}\ }\textbf {\bibinfo {volume} {90}},\ \bibinfo {pages} {084009} (\bibinfo {year} {2014})},\ \Eprint {https://arxiv.org/abs/1407.3086} {arXiv:1407.3086 [gr-qc]} \BibitemShut {NoStop}%
\bibitem [{\citenamefont {Konoplya}\ \emph {et~al.}(2016)\citenamefont {Konoplya}, \citenamefont {Rezzolla},\ and\ \citenamefont {Zhidenko}}]{Konoplya:2016jvv}%
  \BibitemOpen
  \bibfield  {author} {\bibinfo {author} {\bibfnamefont {R.}~\bibnamefont {Konoplya}}, \bibinfo {author} {\bibfnamefont {L.}~\bibnamefont {Rezzolla}},\ and\ \bibinfo {author} {\bibfnamefont {A.}~\bibnamefont {Zhidenko}},\ }\bibfield  {title} {\bibinfo {title} {{General parametrization of axisymmetric black holes in metric theories of gravity}},\ }\href {https://doi.org/10.1103/PhysRevD.93.064015} {\bibfield  {journal} {\bibinfo  {journal} {Phys. Rev. D}\ }\textbf {\bibinfo {volume} {93}},\ \bibinfo {pages} {064015} (\bibinfo {year} {2016})},\ \Eprint {https://arxiv.org/abs/1602.02378} {arXiv:1602.02378 [gr-qc]} \BibitemShut {NoStop}%
\bibitem [{\citenamefont {Visser}\ and\ \citenamefont {Weinfurtner}(2005)}]{Visser:2004zs}%
  \BibitemOpen
  \bibfield  {author} {\bibinfo {author} {\bibfnamefont {M.}~\bibnamefont {Visser}}\ and\ \bibinfo {author} {\bibfnamefont {S.~E.~C.}\ \bibnamefont {Weinfurtner}},\ }\bibfield  {title} {\bibinfo {title} {{Vortex geometry for the equatorial slice of the Kerr black hole}},\ }\href {https://doi.org/10.1088/0264-9381/22/12/011} {\bibfield  {journal} {\bibinfo  {journal} {Class. Quant. Grav.}\ }\textbf {\bibinfo {volume} {22}},\ \bibinfo {pages} {2493} (\bibinfo {year} {2005})},\ \Eprint {https://arxiv.org/abs/gr-qc/0409014} {arXiv:gr-qc/0409014} \BibitemShut {NoStop}%
\bibitem [{\citenamefont {Patrick}\ \emph {et~al.}(2018)\citenamefont {Patrick}, \citenamefont {Coutant}, \citenamefont {Richartz},\ and\ \citenamefont {Weinfurtner}}]{Patrick:2018orp}%
  \BibitemOpen
  \bibfield  {author} {\bibinfo {author} {\bibfnamefont {S.}~\bibnamefont {Patrick}}, \bibinfo {author} {\bibfnamefont {A.}~\bibnamefont {Coutant}}, \bibinfo {author} {\bibfnamefont {M.}~\bibnamefont {Richartz}},\ and\ \bibinfo {author} {\bibfnamefont {S.}~\bibnamefont {Weinfurtner}},\ }\bibfield  {title} {\bibinfo {title} {{Black hole quasibound states from a draining bathtub vortex flow}},\ }\href {https://doi.org/10.1103/PhysRevLett.121.061101} {\bibfield  {journal} {\bibinfo  {journal} {Phys. Rev. Lett.}\ }\textbf {\bibinfo {volume} {121}},\ \bibinfo {pages} {061101} (\bibinfo {year} {2018})},\ \Eprint {https://arxiv.org/abs/1801.08473} {arXiv:1801.08473 [gr-qc]} \BibitemShut {NoStop}%
\bibitem [{\citenamefont {Price}(1972)}]{Price:1971fb}%
  \BibitemOpen
  \bibfield  {author} {\bibinfo {author} {\bibfnamefont {R.~H.}\ \bibnamefont {Price}},\ }\bibfield  {title} {\bibinfo {title} {{Nonspherical perturbations of relativistic gravitational collapse. 1. Scalar and gravitational perturbations}},\ }\href {https://doi.org/10.1103/PhysRevD.5.2419} {\bibfield  {journal} {\bibinfo  {journal} {Phys. Rev. D}\ }\textbf {\bibinfo {volume} {5}},\ \bibinfo {pages} {2419} (\bibinfo {year} {1972})}\BibitemShut {NoStop}%
\bibitem [{\citenamefont {Leaver}(1986)}]{Leaver:1986gd}%
  \BibitemOpen
  \bibfield  {author} {\bibinfo {author} {\bibfnamefont {E.~W.}\ \bibnamefont {Leaver}},\ }\bibfield  {title} {\bibinfo {title} {{Spectral decomposition of the perturbation response of the Schwarzschild geometry}},\ }\href {https://doi.org/10.1103/PhysRevD.34.384} {\bibfield  {journal} {\bibinfo  {journal} {Phys. Rev. D}\ }\textbf {\bibinfo {volume} {34}},\ \bibinfo {pages} {384} (\bibinfo {year} {1986})}\BibitemShut {NoStop}%
\bibitem [{\citenamefont {Buonanno}\ \emph {et~al.}(2007)\citenamefont {Buonanno}, \citenamefont {Cook},\ and\ \citenamefont {Pretorius}}]{Buonanno:2006ui}%
  \BibitemOpen
  \bibfield  {author} {\bibinfo {author} {\bibfnamefont {A.}~\bibnamefont {Buonanno}}, \bibinfo {author} {\bibfnamefont {G.~B.}\ \bibnamefont {Cook}},\ and\ \bibinfo {author} {\bibfnamefont {F.}~\bibnamefont {Pretorius}},\ }\bibfield  {title} {\bibinfo {title} {{Inspiral, merger and ring-down of equal-mass black-hole binaries}},\ }\href {https://doi.org/10.1103/PhysRevD.75.124018} {\bibfield  {journal} {\bibinfo  {journal} {Phys. Rev. D}\ }\textbf {\bibinfo {volume} {75}},\ \bibinfo {pages} {124018} (\bibinfo {year} {2007})},\ \Eprint {https://arxiv.org/abs/gr-qc/0610122} {arXiv:gr-qc/0610122} \BibitemShut {NoStop}%
\bibitem [{\citenamefont {Baibhav}\ \emph {et~al.}(2023)\citenamefont {Baibhav}, \citenamefont {Cheung}, \citenamefont {Berti}, \citenamefont {Cardoso}, \citenamefont {Carullo}, \citenamefont {Cotesta}, \citenamefont {Del~Pozzo},\ and\ \citenamefont {Duque}}]{Baibhav:2023clw}%
  \BibitemOpen
  \bibfield  {author} {\bibinfo {author} {\bibfnamefont {V.}~\bibnamefont {Baibhav}}, \bibinfo {author} {\bibfnamefont {M.~H.-Y.}\ \bibnamefont {Cheung}}, \bibinfo {author} {\bibfnamefont {E.}~\bibnamefont {Berti}}, \bibinfo {author} {\bibfnamefont {V.}~\bibnamefont {Cardoso}}, \bibinfo {author} {\bibfnamefont {G.}~\bibnamefont {Carullo}}, \bibinfo {author} {\bibfnamefont {R.}~\bibnamefont {Cotesta}}, \bibinfo {author} {\bibfnamefont {W.}~\bibnamefont {Del~Pozzo}},\ and\ \bibinfo {author} {\bibfnamefont {F.}~\bibnamefont {Duque}},\ }\bibfield  {title} {\bibinfo {title} {{Agnostic black hole spectroscopy: Quasinormal mode content of numerical relativity waveforms and limits of validity of linear perturbation theory}},\ }\href {https://doi.org/10.1103/PhysRevD.108.104020} {\bibfield  {journal} {\bibinfo  {journal} {Phys. Rev. D}\ }\textbf {\bibinfo {volume} {108}},\ \bibinfo {pages} {104020} (\bibinfo {year} {2023})},\ \Eprint {https://arxiv.org/abs/2302.03050} {arXiv:2302.03050 [gr-qc]} \BibitemShut {NoStop}%
\bibitem [{\citenamefont {Nee}\ \emph {et~al.}(2023)\citenamefont {Nee}, \citenamefont {V\"olkel},\ and\ \citenamefont {Pfeiffer}}]{Nee:2023osy}%
  \BibitemOpen
  \bibfield  {author} {\bibinfo {author} {\bibfnamefont {P.~J.}\ \bibnamefont {Nee}}, \bibinfo {author} {\bibfnamefont {S.~H.}\ \bibnamefont {V\"olkel}},\ and\ \bibinfo {author} {\bibfnamefont {H.~P.}\ \bibnamefont {Pfeiffer}},\ }\bibfield  {title} {\bibinfo {title} {{Role of black hole quasinormal mode overtones for ringdown analysis}},\ }\href {https://doi.org/10.1103/PhysRevD.108.044032} {\bibfield  {journal} {\bibinfo  {journal} {Phys. Rev. D}\ }\textbf {\bibinfo {volume} {108}},\ \bibinfo {pages} {044032} (\bibinfo {year} {2023})},\ \Eprint {https://arxiv.org/abs/2302.06634} {arXiv:2302.06634 [gr-qc]} \BibitemShut {NoStop}%
\bibitem [{\citenamefont {Thomopoulos}\ \emph {et~al.}(2025)\citenamefont {Thomopoulos}, \citenamefont {Völkel},\ and\ \citenamefont {Pfeiffer}}]{thomopoulos2025ringdownspectroscopyphenomenologicallymodified}%
  \BibitemOpen
  \bibfield  {author} {\bibinfo {author} {\bibfnamefont {S.}~\bibnamefont {Thomopoulos}}, \bibinfo {author} {\bibfnamefont {S.~H.}\ \bibnamefont {Völkel}},\ and\ \bibinfo {author} {\bibfnamefont {H.~P.}\ \bibnamefont {Pfeiffer}},\ }\href {https://arxiv.org/abs/2504.17848} {\bibinfo {title} {Ringdown spectroscopy of phenomenologically modified black holes}} (\bibinfo {year} {2025}),\ \Eprint {https://arxiv.org/abs/2504.17848} {arXiv:2504.17848 [gr-qc]} \BibitemShut {NoStop}%
\bibitem [{\citenamefont {Buonanno}\ and\ \citenamefont {Damour}(1999)}]{Buonanno:1998gg}%
  \BibitemOpen
  \bibfield  {author} {\bibinfo {author} {\bibfnamefont {A.}~\bibnamefont {Buonanno}}\ and\ \bibinfo {author} {\bibfnamefont {T.}~\bibnamefont {Damour}},\ }\bibfield  {title} {\bibinfo {title} {{Effective one-body approach to general relativistic two-body dynamics}},\ }\href {https://doi.org/10.1103/PhysRevD.59.084006} {\bibfield  {journal} {\bibinfo  {journal} {Phys. Rev. D}\ }\textbf {\bibinfo {volume} {59}},\ \bibinfo {pages} {084006} (\bibinfo {year} {1999})},\ \Eprint {https://arxiv.org/abs/gr-qc/9811091} {arXiv:gr-qc/9811091} \BibitemShut {NoStop}%
\bibitem [{\citenamefont {Ajith}\ \emph {et~al.}(2011)\citenamefont {Ajith} \emph {et~al.}}]{Ajith:2009bn}%
  \BibitemOpen
  \bibfield  {author} {\bibinfo {author} {\bibfnamefont {P.}~\bibnamefont {Ajith}} \emph {et~al.},\ }\bibfield  {title} {\bibinfo {title} {{Inspiral-merger-ringdown waveforms for black-hole binaries with non-precessing spins}},\ }\href {https://doi.org/10.1103/PhysRevLett.106.241101} {\bibfield  {journal} {\bibinfo  {journal} {Phys. Rev. Lett.}\ }\textbf {\bibinfo {volume} {106}},\ \bibinfo {pages} {241101} (\bibinfo {year} {2011})},\ \Eprint {https://arxiv.org/abs/0909.2867} {arXiv:0909.2867 [gr-qc]} \BibitemShut {NoStop}%
\bibitem [{\citenamefont {Field}\ \emph {et~al.}(2014)\citenamefont {Field}, \citenamefont {Galley}, \citenamefont {Hesthaven}, \citenamefont {Kaye},\ and\ \citenamefont {Tiglio}}]{Field:2013cfa}%
  \BibitemOpen
  \bibfield  {author} {\bibinfo {author} {\bibfnamefont {S.~E.}\ \bibnamefont {Field}}, \bibinfo {author} {\bibfnamefont {C.~R.}\ \bibnamefont {Galley}}, \bibinfo {author} {\bibfnamefont {J.~S.}\ \bibnamefont {Hesthaven}}, \bibinfo {author} {\bibfnamefont {J.}~\bibnamefont {Kaye}},\ and\ \bibinfo {author} {\bibfnamefont {M.}~\bibnamefont {Tiglio}},\ }\bibfield  {title} {\bibinfo {title} {{Fast prediction and evaluation of gravitational waveforms using surrogate models}},\ }\href {https://doi.org/10.1103/PhysRevX.4.031006} {\bibfield  {journal} {\bibinfo  {journal} {Phys. Rev. X}\ }\textbf {\bibinfo {volume} {4}},\ \bibinfo {pages} {031006} (\bibinfo {year} {2014})},\ \Eprint {https://arxiv.org/abs/1308.3565} {arXiv:1308.3565 [gr-qc]} \BibitemShut {NoStop}%
\bibitem [{\citenamefont {P\"urrer}(2014)}]{Purrer:2014fza}%
  \BibitemOpen
  \bibfield  {author} {\bibinfo {author} {\bibfnamefont {M.}~\bibnamefont {P\"urrer}},\ }\bibfield  {title} {\bibinfo {title} {{Frequency domain reduced order models for gravitational waves from aligned-spin compact binaries}},\ }\href {https://doi.org/10.1088/0264-9381/31/19/195010} {\bibfield  {journal} {\bibinfo  {journal} {Class. Quant. Grav.}\ }\textbf {\bibinfo {volume} {31}},\ \bibinfo {pages} {195010} (\bibinfo {year} {2014})},\ \Eprint {https://arxiv.org/abs/1402.4146} {arXiv:1402.4146 [gr-qc]} \BibitemShut {NoStop}%
\bibitem [{\citenamefont {Maggiore}(2007)}]{Maggiore:2007ulw}%
  \BibitemOpen
  \bibfield  {author} {\bibinfo {author} {\bibfnamefont {M.}~\bibnamefont {Maggiore}},\ }\href {https://doi.org/10.1093/acprof:oso/9780198570745.001.0001} {\emph {\bibinfo {title} {{Gravitational Waves. Vol. 1: Theory and Experiments}}}}\ (\bibinfo  {publisher} {Oxford University Press},\ \bibinfo {year} {2007})\BibitemShut {NoStop}%
\bibitem [{\citenamefont {Maggiore}(2018)}]{Maggiore:2018sht}%
  \BibitemOpen
  \bibfield  {author} {\bibinfo {author} {\bibfnamefont {M.}~\bibnamefont {Maggiore}},\ }\href@noop {} {\emph {\bibinfo {title} {{Gravitational Waves. Vol. 2: Astrophysics and Cosmology}}}}\ (\bibinfo  {publisher} {Oxford University Press},\ \bibinfo {year} {2018})\BibitemShut {NoStop}%
\bibitem [{\citenamefont {Chatziioannou}\ \emph {et~al.}(2024)\citenamefont {Chatziioannou}, \citenamefont {Dent}, \citenamefont {Fishbach}, \citenamefont {Ohme}, \citenamefont {Pürrer}, \citenamefont {Raymond},\ and\ \citenamefont {Veitch}}]{chatziioannou2024compactbinarycoalescencesgravitationalwave}%
  \BibitemOpen
  \bibfield  {author} {\bibinfo {author} {\bibfnamefont {K.}~\bibnamefont {Chatziioannou}}, \bibinfo {author} {\bibfnamefont {T.}~\bibnamefont {Dent}}, \bibinfo {author} {\bibfnamefont {M.}~\bibnamefont {Fishbach}}, \bibinfo {author} {\bibfnamefont {F.}~\bibnamefont {Ohme}}, \bibinfo {author} {\bibfnamefont {M.}~\bibnamefont {Pürrer}}, \bibinfo {author} {\bibfnamefont {V.}~\bibnamefont {Raymond}},\ and\ \bibinfo {author} {\bibfnamefont {J.}~\bibnamefont {Veitch}},\ }\href {https://arxiv.org/abs/2409.02037} {\bibinfo {title} {Compact binary coalescences: gravitational-wave astronomy with ground-based detectors}} (\bibinfo {year} {2024}),\ \Eprint {https://arxiv.org/abs/2409.02037} {arXiv:2409.02037 [gr-qc]} \BibitemShut {NoStop}%
\bibitem [{\citenamefont {Johannsen}\ and\ \citenamefont {Psaltis}(2011)}]{Johannsen:2011dh}%
  \BibitemOpen
  \bibfield  {author} {\bibinfo {author} {\bibfnamefont {T.}~\bibnamefont {Johannsen}}\ and\ \bibinfo {author} {\bibfnamefont {D.}~\bibnamefont {Psaltis}},\ }\bibfield  {title} {\bibinfo {title} {{A Metric for Rapidly Spinning Black Holes Suitable for Strong-Field Tests of the No-Hair Theorem}},\ }\href {https://doi.org/10.1103/PhysRevD.83.124015} {\bibfield  {journal} {\bibinfo  {journal} {Phys. Rev. D}\ }\textbf {\bibinfo {volume} {83}},\ \bibinfo {pages} {124015} (\bibinfo {year} {2011})},\ \Eprint {https://arxiv.org/abs/1105.3191} {arXiv:1105.3191 [gr-qc]} \BibitemShut {NoStop}%
\bibitem [{\citenamefont {Johannsen}(2013)}]{Johannsen:2013szh}%
  \BibitemOpen
  \bibfield  {author} {\bibinfo {author} {\bibfnamefont {T.}~\bibnamefont {Johannsen}},\ }\bibfield  {title} {\bibinfo {title} {{Regular Black Hole Metric with Three Constants of Motion}},\ }\href {https://doi.org/10.1103/PhysRevD.88.044002} {\bibfield  {journal} {\bibinfo  {journal} {Phys. Rev. D}\ }\textbf {\bibinfo {volume} {88}},\ \bibinfo {pages} {044002} (\bibinfo {year} {2013})},\ \Eprint {https://arxiv.org/abs/1501.02809} {arXiv:1501.02809 [gr-qc]} \BibitemShut {NoStop}%
\bibitem [{\citenamefont {Konoplya}\ \emph {et~al.}(2020)\citenamefont {Konoplya}, \citenamefont {Pappas},\ and\ \citenamefont {Stuchl\'\i{}k}}]{Konoplya:2020kqb}%
  \BibitemOpen
  \bibfield  {author} {\bibinfo {author} {\bibfnamefont {R.~A.}\ \bibnamefont {Konoplya}}, \bibinfo {author} {\bibfnamefont {T.~D.}\ \bibnamefont {Pappas}},\ and\ \bibinfo {author} {\bibfnamefont {Z.}~\bibnamefont {Stuchl\'\i{}k}},\ }\bibfield  {title} {\bibinfo {title} {{General parametrization of higher-dimensional black holes and its application to Einstein-Lovelock theory}},\ }\href {https://doi.org/10.1103/PhysRevD.102.084043} {\bibfield  {journal} {\bibinfo  {journal} {Phys. Rev. D}\ }\textbf {\bibinfo {volume} {102}},\ \bibinfo {pages} {084043} (\bibinfo {year} {2020})},\ \Eprint {https://arxiv.org/abs/2007.14860} {arXiv:2007.14860 [gr-qc]} \BibitemShut {NoStop}%
\bibitem [{\citenamefont {Konoplya}\ and\ \citenamefont {Zhidenko}(2020)}]{Konoplya:2020hyk}%
  \BibitemOpen
  \bibfield  {author} {\bibinfo {author} {\bibfnamefont {R.~A.}\ \bibnamefont {Konoplya}}\ and\ \bibinfo {author} {\bibfnamefont {A.}~\bibnamefont {Zhidenko}},\ }\bibfield  {title} {\bibinfo {title} {{General parametrization of black holes: The only parameters that matter}},\ }\href {https://doi.org/10.1103/PhysRevD.101.124004} {\bibfield  {journal} {\bibinfo  {journal} {Phys. Rev. D}\ }\textbf {\bibinfo {volume} {101}},\ \bibinfo {pages} {124004} (\bibinfo {year} {2020})},\ \Eprint {https://arxiv.org/abs/2001.06100} {arXiv:2001.06100 [gr-qc]} \BibitemShut {NoStop}%
\bibitem [{\citenamefont {Schutzhold}\ and\ \citenamefont {Unruh}(2002)}]{Schutzhold:2002rf}%
  \BibitemOpen
  \bibfield  {author} {\bibinfo {author} {\bibfnamefont {R.}~\bibnamefont {Schutzhold}}\ and\ \bibinfo {author} {\bibfnamefont {W.~G.}\ \bibnamefont {Unruh}},\ }\bibfield  {title} {\bibinfo {title} {{Gravity wave analogs of black holes}},\ }\href {https://doi.org/10.1103/PhysRevD.66.044019} {\bibfield  {journal} {\bibinfo  {journal} {Phys. Rev. D}\ }\textbf {\bibinfo {volume} {66}},\ \bibinfo {pages} {044019} (\bibinfo {year} {2002})},\ \Eprint {https://arxiv.org/abs/gr-qc/0205099} {arXiv:gr-qc/0205099} \BibitemShut {NoStop}%
\bibitem [{\citenamefont {Dolan}\ \emph {et~al.}(2012)\citenamefont {Dolan}, \citenamefont {Oliveira},\ and\ \citenamefont {Crispino}}]{Dolan:2011ti}%
  \BibitemOpen
  \bibfield  {author} {\bibinfo {author} {\bibfnamefont {S.~R.}\ \bibnamefont {Dolan}}, \bibinfo {author} {\bibfnamefont {L.~A.}\ \bibnamefont {Oliveira}},\ and\ \bibinfo {author} {\bibfnamefont {L.~C.~B.}\ \bibnamefont {Crispino}},\ }\bibfield  {title} {\bibinfo {title} {{Resonances of a rotating black hole analogue}},\ }\href {https://doi.org/10.1103/PhysRevD.85.044031} {\bibfield  {journal} {\bibinfo  {journal} {Phys. Rev. D}\ }\textbf {\bibinfo {volume} {85}},\ \bibinfo {pages} {044031} (\bibinfo {year} {2012})},\ \Eprint {https://arxiv.org/abs/1105.1795} {arXiv:1105.1795 [gr-qc]} \BibitemShut {NoStop}%
\bibitem [{\citenamefont {Torres}\ \emph {et~al.}(2022)\citenamefont {Torres}, \citenamefont {Patrick},\ and\ \citenamefont {Gregory}}]{Torres:2022bto}%
  \BibitemOpen
  \bibfield  {author} {\bibinfo {author} {\bibfnamefont {T.}~\bibnamefont {Torres}}, \bibinfo {author} {\bibfnamefont {S.}~\bibnamefont {Patrick}},\ and\ \bibinfo {author} {\bibfnamefont {R.}~\bibnamefont {Gregory}},\ }\bibfield  {title} {\bibinfo {title} {{Imperfect draining vortex as analog extreme compact object}},\ }\href {https://doi.org/10.1103/PhysRevD.106.045026} {\bibfield  {journal} {\bibinfo  {journal} {Phys. Rev. D}\ }\textbf {\bibinfo {volume} {106}},\ \bibinfo {pages} {045026} (\bibinfo {year} {2022})},\ \Eprint {https://arxiv.org/abs/2204.10139} {arXiv:2204.10139 [gr-qc]} \BibitemShut {NoStop}%
\bibitem [{\citenamefont {Sivia}\ and\ \citenamefont {Skilling}(2006)}]{sivia2006data}%
  \BibitemOpen
  \bibfield  {author} {\bibinfo {author} {\bibfnamefont {D.}~\bibnamefont {Sivia}}\ and\ \bibinfo {author} {\bibfnamefont {J.}~\bibnamefont {Skilling}},\ }\href {https://books.google.de/books?id=lYMSDAAAQBAJ} {\emph {\bibinfo {title} {Data Analysis: A Bayesian Tutorial}}},\ Oxford science publications\ (\bibinfo  {publisher} {OUP Oxford},\ \bibinfo {year} {2006})\BibitemShut {NoStop}%
\bibitem [{\citenamefont {Metropolis}\ \emph {et~al.}(1953)\citenamefont {Metropolis}, \citenamefont {Rosenbluth}, \citenamefont {Rosenbluth}, \citenamefont {Teller},\ and\ \citenamefont {Teller}}]{Metropolis:1953am}%
  \BibitemOpen
  \bibfield  {author} {\bibinfo {author} {\bibfnamefont {N.}~\bibnamefont {Metropolis}}, \bibinfo {author} {\bibfnamefont {A.~W.}\ \bibnamefont {Rosenbluth}}, \bibinfo {author} {\bibfnamefont {M.~N.}\ \bibnamefont {Rosenbluth}}, \bibinfo {author} {\bibfnamefont {A.~H.}\ \bibnamefont {Teller}},\ and\ \bibinfo {author} {\bibfnamefont {E.}~\bibnamefont {Teller}},\ }\bibfield  {title} {\bibinfo {title} {{Equation of state calculations by fast computing machines}},\ }\href {https://doi.org/10.1063/1.1699114} {\bibfield  {journal} {\bibinfo  {journal} {J. Chem. Phys.}\ }\textbf {\bibinfo {volume} {21}},\ \bibinfo {pages} {1087} (\bibinfo {year} {1953})}\BibitemShut {NoStop}%
\bibitem [{\citenamefont {Foreman-Mackey}\ \emph {et~al.}(2013)\citenamefont {Foreman-Mackey}, \citenamefont {Hogg}, \citenamefont {Lang},\ and\ \citenamefont {Goodman}}]{Foreman-Mackey:2012any}%
  \BibitemOpen
  \bibfield  {author} {\bibinfo {author} {\bibfnamefont {D.}~\bibnamefont {Foreman-Mackey}}, \bibinfo {author} {\bibfnamefont {D.~W.}\ \bibnamefont {Hogg}}, \bibinfo {author} {\bibfnamefont {D.}~\bibnamefont {Lang}},\ and\ \bibinfo {author} {\bibfnamefont {J.}~\bibnamefont {Goodman}},\ }\bibfield  {title} {\bibinfo {title} {{emcee: The MCMC Hammer}},\ }\href {https://doi.org/10.1086/670067} {\bibfield  {journal} {\bibinfo  {journal} {Publ. Astron. Soc. Pac.}\ }\textbf {\bibinfo {volume} {125}},\ \bibinfo {pages} {306} (\bibinfo {year} {2013})},\ \Eprint {https://arxiv.org/abs/1202.3665} {arXiv:1202.3665 [astro-ph.IM]} \BibitemShut {NoStop}%
\bibitem [{\citenamefont {{Goodman}}\ and\ \citenamefont {{Weare}}(2010)}]{2010CAMCS...5...65G}%
  \BibitemOpen
  \bibfield  {author} {\bibinfo {author} {\bibfnamefont {J.}~\bibnamefont {{Goodman}}}\ and\ \bibinfo {author} {\bibfnamefont {J.}~\bibnamefont {{Weare}}},\ }\bibfield  {title} {\bibinfo {title} {{Ensemble samplers with affine invariance}},\ }\href {https://doi.org/10.2140/camcos.2010.5.65} {\bibfield  {journal} {\bibinfo  {journal} {Communications in Applied Mathematics and Computational Science}\ }\textbf {\bibinfo {volume} {5}},\ \bibinfo {pages} {65} (\bibinfo {year} {2010})}\BibitemShut {NoStop}%
\bibitem [{\citenamefont {Kocherlakota}\ and\ \citenamefont {Rezzolla}(2022{\natexlab{a}})}]{Kocherlakota:2022jnz}%
  \BibitemOpen
  \bibfield  {author} {\bibinfo {author} {\bibfnamefont {P.}~\bibnamefont {Kocherlakota}}\ and\ \bibinfo {author} {\bibfnamefont {L.}~\bibnamefont {Rezzolla}},\ }\bibfield  {title} {\bibinfo {title} {{Distinguishing gravitational and emission physics in black hole imaging: spherical symmetry}},\ }\href {https://doi.org/10.1093/mnras/stac891} {\bibfield  {journal} {\bibinfo  {journal} {Mon. Not. Roy. Astron. Soc.}\ }\textbf {\bibinfo {volume} {513}},\ \bibinfo {pages} {1229} (\bibinfo {year} {2022}{\natexlab{a}})},\ \Eprint {https://arxiv.org/abs/2201.05641} {arXiv:2201.05641 [gr-qc]} \BibitemShut {NoStop}%
\bibitem [{\citenamefont {Kocherlakota}\ and\ \citenamefont {Rezzolla}(2022{\natexlab{b}})}]{kocherlakota2022commentanalyticalboundsrezzollazhidenko}%
  \BibitemOpen
  \bibfield  {author} {\bibinfo {author} {\bibfnamefont {P.}~\bibnamefont {Kocherlakota}}\ and\ \bibinfo {author} {\bibfnamefont {L.}~\bibnamefont {Rezzolla}},\ }\href {https://arxiv.org/abs/2206.03146} {\bibinfo {title} {Comment on the analytical bounds in the rezzolla-zhidenko parametrization}} (\bibinfo {year} {2022}{\natexlab{b}}),\ \Eprint {https://arxiv.org/abs/2206.03146} {arXiv:2206.03146 [gr-qc]} \BibitemShut {NoStop}%
\bibitem [{\citenamefont {Matyjasek}\ \emph {et~al.}(2024)\citenamefont {Matyjasek}, \citenamefont {Benda},\ and\ \citenamefont {Stafi\'nska}}]{Matyjasek:2024uwo}%
  \BibitemOpen
  \bibfield  {author} {\bibinfo {author} {\bibfnamefont {J.}~\bibnamefont {Matyjasek}}, \bibinfo {author} {\bibfnamefont {K.}~\bibnamefont {Benda}},\ and\ \bibinfo {author} {\bibfnamefont {M.}~\bibnamefont {Stafi\'nska}},\ }\bibfield  {title} {\bibinfo {title} {{Accurate quasinormal modes of the analog black holes}},\ }\href {https://doi.org/10.1103/PhysRevD.110.064083} {\bibfield  {journal} {\bibinfo  {journal} {Phys. Rev. D}\ }\textbf {\bibinfo {volume} {110}},\ \bibinfo {pages} {064083} (\bibinfo {year} {2024})},\ \Eprint {https://arxiv.org/abs/2408.16116} {arXiv:2408.16116 [gr-qc]} \BibitemShut {NoStop}%
\bibitem [{\citenamefont {Wheeler}(2015)}]{lieb2015studies}%
  \BibitemOpen
  \bibfield  {author} {\bibinfo {author} {\bibfnamefont {J.~A.}\ \bibnamefont {Wheeler}},\ }\href {https://press.princeton.edu/titles/861.html} {\emph {\bibinfo {title} {{Studies in Mathematical Physics: Essays in Honor of Valentine Bargmann}}}},\ Princeton Series in Physics\ (\bibinfo  {publisher} {Princeton University Press},\ \bibinfo {year} {2015})\ pp.\ \bibinfo {pages} {351--422}\BibitemShut {NoStop}%
\bibitem [{\citenamefont {Chadan}\ and\ \citenamefont {Sabatier}(1989)}]{MR985100}%
  \BibitemOpen
  \bibfield  {author} {\bibinfo {author} {\bibfnamefont {K.}~\bibnamefont {Chadan}}\ and\ \bibinfo {author} {\bibfnamefont {P.~C.}\ \bibnamefont {Sabatier}},\ }\href {https://doi.org/10.1007/978-3-642-83317-5} {\emph {\bibinfo {title} {{Inverse problems in quantum scattering theory}}}},\ \bibinfo {edition} {2nd}\ ed.,\ Texts and Monographs in Physics\ (\bibinfo  {publisher} {Springer-Verlag},\ \bibinfo {address} {New York},\ \bibinfo {year} {1989})\BibitemShut {NoStop}%
\bibitem [{\citenamefont {{Lazenby}}\ and\ \citenamefont {{Griffiths}}(1980)}]{1980AmJPh..48..432L}%
  \BibitemOpen
  \bibfield  {author} {\bibinfo {author} {\bibfnamefont {J.~C.}\ \bibnamefont {{Lazenby}}}\ and\ \bibinfo {author} {\bibfnamefont {D.~J.}\ \bibnamefont {{Griffiths}}},\ }\bibfield  {title} {\bibinfo {title} {{Classical inverse scattering in one dimension}},\ }\href {https://doi.org/10.1119/1.11998} {\bibfield  {journal} {\bibinfo  {journal} {American Journal of Physics}\ }\textbf {\bibinfo {volume} {48}},\ \bibinfo {pages} {432} (\bibinfo {year} {1980})}\BibitemShut {NoStop}%
\bibitem [{\citenamefont {{Gandhi}}\ and\ \citenamefont {{Efthimiou}}(2006)}]{2006AmJPh..74..638G}%
  \BibitemOpen
  \bibfield  {author} {\bibinfo {author} {\bibfnamefont {S.~C.}\ \bibnamefont {{Gandhi}}}\ and\ \bibinfo {author} {\bibfnamefont {C.~J.}\ \bibnamefont {{Efthimiou}}},\ }\bibfield  {title} {\bibinfo {title} {{Inversion of Gamow's formula and inverse scattering}},\ }\href {https://doi.org/10.1119/1.2190683} {\bibfield  {journal} {\bibinfo  {journal} {American Journal of Physics}\ }\textbf {\bibinfo {volume} {74}},\ \bibinfo {pages} {638} (\bibinfo {year} {2006})},\ \Eprint {https://arxiv.org/abs/quant-ph/0503223} {quant-ph/0503223} \BibitemShut {NoStop}%
\bibitem [{\citenamefont {Albuquerque}\ \emph {et~al.}(2023)\citenamefont {Albuquerque}, \citenamefont {V\"olkel}, \citenamefont {Kokkotas},\ and\ \citenamefont {Bezerra}}]{Albuquerque:2023lzw}%
  \BibitemOpen
  \bibfield  {author} {\bibinfo {author} {\bibfnamefont {S.}~\bibnamefont {Albuquerque}}, \bibinfo {author} {\bibfnamefont {S.~H.}\ \bibnamefont {V\"olkel}}, \bibinfo {author} {\bibfnamefont {K.~D.}\ \bibnamefont {Kokkotas}},\ and\ \bibinfo {author} {\bibfnamefont {V.~B.}\ \bibnamefont {Bezerra}},\ }\bibfield  {title} {\bibinfo {title} {{Inverse problem of analog gravity systems}},\ }\href {https://doi.org/10.1103/PhysRevD.108.124053} {\bibfield  {journal} {\bibinfo  {journal} {Phys. Rev. D}\ }\textbf {\bibinfo {volume} {108}},\ \bibinfo {pages} {124053} (\bibinfo {year} {2023})},\ \Eprint {https://arxiv.org/abs/2309.11168} {arXiv:2309.11168 [gr-qc]} \BibitemShut {NoStop}%
\bibitem [{\citenamefont {Albuquerque}\ \emph {et~al.}(2024{\natexlab{a}})\citenamefont {Albuquerque}, \citenamefont {V\"olkel},\ and\ \citenamefont {Kokkotas}}]{Albuquerque:2024xol}%
  \BibitemOpen
  \bibfield  {author} {\bibinfo {author} {\bibfnamefont {S.}~\bibnamefont {Albuquerque}}, \bibinfo {author} {\bibfnamefont {S.~H.}\ \bibnamefont {V\"olkel}},\ and\ \bibinfo {author} {\bibfnamefont {K.~D.}\ \bibnamefont {Kokkotas}},\ }\bibfield  {title} {\bibinfo {title} {{Inverse problem in energy-dependent potentials using semiclassical methods}},\ }\href {https://doi.org/10.1103/PhysRevD.109.096014} {\bibfield  {journal} {\bibinfo  {journal} {Phys. Rev. D}\ }\textbf {\bibinfo {volume} {109}},\ \bibinfo {pages} {096014} (\bibinfo {year} {2024}{\natexlab{a}})},\ \Eprint {https://arxiv.org/abs/2404.11478} {arXiv:2404.11478 [hep-ph]} \BibitemShut {NoStop}%
\bibitem [{\citenamefont {Albuquerque}\ \emph {et~al.}(2024{\natexlab{b}})\citenamefont {Albuquerque}, \citenamefont {V\"olkel}, \citenamefont {Kokkotas},\ and\ \citenamefont {Bezerra}}]{Albuquerque:2024cwl}%
  \BibitemOpen
  \bibfield  {author} {\bibinfo {author} {\bibfnamefont {S.}~\bibnamefont {Albuquerque}}, \bibinfo {author} {\bibfnamefont {S.~H.}\ \bibnamefont {V\"olkel}}, \bibinfo {author} {\bibfnamefont {K.~D.}\ \bibnamefont {Kokkotas}},\ and\ \bibinfo {author} {\bibfnamefont {V.~B.}\ \bibnamefont {Bezerra}},\ }\bibfield  {title} {\bibinfo {title} {{Inverse problem of analog gravity systems. II. Rotation and energy-dependent boundary conditions}},\ }\href {https://doi.org/10.1103/PhysRevD.110.064084} {\bibfield  {journal} {\bibinfo  {journal} {Phys. Rev. D}\ }\textbf {\bibinfo {volume} {110}},\ \bibinfo {pages} {064084} (\bibinfo {year} {2024}{\natexlab{b}})},\ \Eprint {https://arxiv.org/abs/2406.16670} {arXiv:2406.16670 [gr-qc]} \BibitemShut {NoStop}%
\bibitem [{\citenamefont {Chandrasekhar}\ and\ \citenamefont {Ferrari}(1991)}]{doi:10.1098/rspa.1991.0104}%
  \BibitemOpen
  \bibfield  {author} {\bibinfo {author} {\bibfnamefont {S.}~\bibnamefont {Chandrasekhar}}\ and\ \bibinfo {author} {\bibfnamefont {V.}~\bibnamefont {Ferrari}},\ }\bibfield  {title} {\bibinfo {title} {{On the non-radial oscillations of a star. III. A reconsideration of the axial modes}},\ }\href {https://doi.org/10.1098/rspa.1991.0104} {\bibfield  {journal} {\bibinfo  {journal} {Proceedings of the Royal Society of London. Series A: Mathematical and Physical Sciences}\ }\textbf {\bibinfo {volume} {434}},\ \bibinfo {pages} {449} (\bibinfo {year} {1991})}\BibitemShut {NoStop}%
\bibitem [{\citenamefont {Kokkotas}(1994)}]{Kokkotas:1994an}%
  \BibitemOpen
  \bibfield  {author} {\bibinfo {author} {\bibfnamefont {K.~D.}\ \bibnamefont {Kokkotas}},\ }\bibfield  {title} {\bibinfo {title} {{Axial modes for relativistic stars}},\ }\href@noop {} {\bibfield  {journal} {\bibinfo  {journal} {Mon. Not. Roy. Astron. Soc.}\ }\textbf {\bibinfo {volume} {268}},\ \bibinfo {pages} {1015} (\bibinfo {year} {1994})}\BibitemShut {NoStop}%
\bibitem [{\citenamefont {V\"olkel}\ and\ \citenamefont {Kokkotas}(2017)}]{Volkel:2017kfj}%
  \BibitemOpen
  \bibfield  {author} {\bibinfo {author} {\bibfnamefont {S.~H.}\ \bibnamefont {V\"olkel}}\ and\ \bibinfo {author} {\bibfnamefont {K.~D.}\ \bibnamefont {Kokkotas}},\ }\bibfield  {title} {\bibinfo {title} {{Ultra Compact Stars: Reconstructing the Perturbation Potential}},\ }\href {https://doi.org/10.1088/1361-6382/aa82de} {\bibfield  {journal} {\bibinfo  {journal} {Class. Quant. Grav.}\ }\textbf {\bibinfo {volume} {34}},\ \bibinfo {pages} {175015} (\bibinfo {year} {2017})},\ \Eprint {https://arxiv.org/abs/1704.07517} {arXiv:1704.07517 [gr-qc]} \BibitemShut {NoStop}%
\bibitem [{\citenamefont {V\"olkel}\ and\ \citenamefont {Kokkotas}(2018)}]{Volkel:2018hwb}%
  \BibitemOpen
  \bibfield  {author} {\bibinfo {author} {\bibfnamefont {S.~H.}\ \bibnamefont {V\"olkel}}\ and\ \bibinfo {author} {\bibfnamefont {K.~D.}\ \bibnamefont {Kokkotas}},\ }\bibfield  {title} {\bibinfo {title} {{Wormhole Potentials and Throats from Quasi-Normal Modes}},\ }\href {https://doi.org/10.1088/1361-6382/aabce6} {\bibfield  {journal} {\bibinfo  {journal} {Class. Quant. Grav.}\ }\textbf {\bibinfo {volume} {35}},\ \bibinfo {pages} {105018} (\bibinfo {year} {2018})},\ \Eprint {https://arxiv.org/abs/1802.08525} {arXiv:1802.08525 [gr-qc]} \BibitemShut {NoStop}%
\bibitem [{\citenamefont {V\"olkel}\ \emph {et~al.}(2019)\citenamefont {V\"olkel}, \citenamefont {Konoplya},\ and\ \citenamefont {Kokkotas}}]{Volkel:2019ahb}%
  \BibitemOpen
  \bibfield  {author} {\bibinfo {author} {\bibfnamefont {S.~H.}\ \bibnamefont {V\"olkel}}, \bibinfo {author} {\bibfnamefont {R.}~\bibnamefont {Konoplya}},\ and\ \bibinfo {author} {\bibfnamefont {K.~D.}\ \bibnamefont {Kokkotas}},\ }\bibfield  {title} {\bibinfo {title} {{Inverse problem for Hawking radiation}},\ }\href {https://doi.org/10.1103/PhysRevD.99.104025} {\bibfield  {journal} {\bibinfo  {journal} {Phys. Rev. D}\ }\textbf {\bibinfo {volume} {99}},\ \bibinfo {pages} {104025} (\bibinfo {year} {2019})},\ \Eprint {https://arxiv.org/abs/1902.07611} {arXiv:1902.07611 [gr-qc]} \BibitemShut {NoStop}%
\bibitem [{\citenamefont {V\"olkel}\ and\ \citenamefont {Barausse}(2020)}]{Volkel:2020daa}%
  \BibitemOpen
  \bibfield  {author} {\bibinfo {author} {\bibfnamefont {S.~H.}\ \bibnamefont {V\"olkel}}\ and\ \bibinfo {author} {\bibfnamefont {E.}~\bibnamefont {Barausse}},\ }\bibfield  {title} {\bibinfo {title} {{Bayesian Metric Reconstruction with Gravitational Wave Observations}},\ }\href {https://doi.org/10.1103/PhysRevD.102.084025} {\bibfield  {journal} {\bibinfo  {journal} {Phys. Rev. D}\ }\textbf {\bibinfo {volume} {102}},\ \bibinfo {pages} {084025} (\bibinfo {year} {2020})},\ \Eprint {https://arxiv.org/abs/2007.02986} {arXiv:2007.02986 [gr-qc]} \BibitemShut {NoStop}%
\bibitem [{\citenamefont {V\"olkel}\ and\ \citenamefont {Kokkotas}(2019)}]{Volkel:2019muj}%
  \BibitemOpen
  \bibfield  {author} {\bibinfo {author} {\bibfnamefont {S.~H.}\ \bibnamefont {V\"olkel}}\ and\ \bibinfo {author} {\bibfnamefont {K.~D.}\ \bibnamefont {Kokkotas}},\ }\bibfield  {title} {\bibinfo {title} {{Scalar Fields and Parametrized Spherically Symmetric Black Holes: Can one hear the shape of space-time?}},\ }\href {https://doi.org/10.1103/PhysRevD.100.044026} {\bibfield  {journal} {\bibinfo  {journal} {Phys. Rev. D}\ }\textbf {\bibinfo {volume} {100}},\ \bibinfo {pages} {044026} (\bibinfo {year} {2019})},\ \Eprint {https://arxiv.org/abs/1908.00252} {arXiv:1908.00252 [gr-qc]} \BibitemShut {NoStop}%
\bibitem [{\citenamefont {Zhu}\ \emph {et~al.}(2024)\citenamefont {Zhu}, \citenamefont {Ripley}, \citenamefont {C\'ardenas-Avenda\~no},\ and\ \citenamefont {Pretorius}}]{Zhu:2023mzv}%
  \BibitemOpen
  \bibfield  {author} {\bibinfo {author} {\bibfnamefont {H.}~\bibnamefont {Zhu}}, \bibinfo {author} {\bibfnamefont {J.~L.}\ \bibnamefont {Ripley}}, \bibinfo {author} {\bibfnamefont {A.}~\bibnamefont {C\'ardenas-Avenda\~no}},\ and\ \bibinfo {author} {\bibfnamefont {F.}~\bibnamefont {Pretorius}},\ }\bibfield  {title} {\bibinfo {title} {{Challenges in quasinormal mode extraction: Perspectives from numerical solutions to the Teukolsky equation}},\ }\href {https://doi.org/10.1103/PhysRevD.109.044010} {\bibfield  {journal} {\bibinfo  {journal} {Phys. Rev. D}\ }\textbf {\bibinfo {volume} {109}},\ \bibinfo {pages} {044010} (\bibinfo {year} {2024})},\ \Eprint {https://arxiv.org/abs/2309.13204} {arXiv:2309.13204 [gr-qc]} \BibitemShut {NoStop}%
\bibitem [{\citenamefont {Sinatra}\ \emph {et~al.}(2002)\citenamefont {Sinatra}, \citenamefont {Lobo},\ and\ \citenamefont {Castin}}]{Sinatra_2002}%
  \BibitemOpen
  \bibfield  {author} {\bibinfo {author} {\bibfnamefont {A.}~\bibnamefont {Sinatra}}, \bibinfo {author} {\bibfnamefont {C.}~\bibnamefont {Lobo}},\ and\ \bibinfo {author} {\bibfnamefont {Y.}~\bibnamefont {Castin}},\ }\bibfield  {title} {\bibinfo {title} {The truncated wigner method for bose-condensed gases: limits of validity and applications},\ }\href {https://doi.org/10.1088/0953-4075/35/17/301} {\bibfield  {journal} {\bibinfo  {journal} {Journal of Physics B: Atomic, Molecular and Optical Physics}\ }\textbf {\bibinfo {volume} {35}},\ \bibinfo {pages} {3599–3631} (\bibinfo {year} {2002})}\BibitemShut {NoStop}%
\bibitem [{\citenamefont {Carusotto}\ \emph {et~al.}(2008)\citenamefont {Carusotto}, \citenamefont {Fagnocchi}, \citenamefont {Recati}, \citenamefont {Balbinot},\ and\ \citenamefont {Fabbri}}]{Carusotto:2008ep}%
  \BibitemOpen
  \bibfield  {author} {\bibinfo {author} {\bibfnamefont {I.}~\bibnamefont {Carusotto}}, \bibinfo {author} {\bibfnamefont {S.}~\bibnamefont {Fagnocchi}}, \bibinfo {author} {\bibfnamefont {A.}~\bibnamefont {Recati}}, \bibinfo {author} {\bibfnamefont {R.}~\bibnamefont {Balbinot}},\ and\ \bibinfo {author} {\bibfnamefont {A.}~\bibnamefont {Fabbri}},\ }\bibfield  {title} {\bibinfo {title} {{Numerical observation of Hawking radiation from acoustic black holes in atomic BECs}},\ }\href {https://doi.org/10.1088/1367-2630/10/10/103001} {\bibfield  {journal} {\bibinfo  {journal} {New J. Phys.}\ }\textbf {\bibinfo {volume} {10}},\ \bibinfo {pages} {103001} (\bibinfo {year} {2008})},\ \Eprint {https://arxiv.org/abs/0803.0507} {arXiv:0803.0507 [cond-mat.other]} \BibitemShut {NoStop}%
\bibitem [{\citenamefont {Jacquet}\ \emph {et~al.}(2022)\citenamefont {Jacquet}, \citenamefont {Joly}, \citenamefont {Giacomelli}, \citenamefont {Claude}, \citenamefont {Glorieux}, \citenamefont {Bramati}, \citenamefont {Carusotto},\ and\ \citenamefont {Giacobino}}]{Jacquet:2022vak}%
  \BibitemOpen
  \bibfield  {author} {\bibinfo {author} {\bibfnamefont {M.~J.}\ \bibnamefont {Jacquet}}, \bibinfo {author} {\bibfnamefont {M.}~\bibnamefont {Joly}}, \bibinfo {author} {\bibfnamefont {L.}~\bibnamefont {Giacomelli}}, \bibinfo {author} {\bibfnamefont {F.}~\bibnamefont {Claude}}, \bibinfo {author} {\bibfnamefont {Q.}~\bibnamefont {Glorieux}}, \bibinfo {author} {\bibfnamefont {A.}~\bibnamefont {Bramati}}, \bibinfo {author} {\bibfnamefont {I.}~\bibnamefont {Carusotto}},\ and\ \bibinfo {author} {\bibfnamefont {E.}~\bibnamefont {Giacobino}},\ }\bibfield  {title} {\bibinfo {title} {{Analogue quantum simulation of the Hawking effect in a polariton superfluid}},\ }\href {https://doi.org/10.1140/epjd/s10053-022-00477-5} {\bibfield  {journal} {\bibinfo  {journal} {Eur. Phys. J. D}\ }\textbf {\bibinfo {volume} {76}},\ \bibinfo {pages} {152} (\bibinfo {year} {2022})},\ \bibinfo {note} {[Erratum: Eur.Phys.J.D 76, 247 (2022)]},\ \Eprint {https://arxiv.org/abs/2201.02038} {arXiv:2201.02038 [quant-ph]} \BibitemShut {NoStop}%
\bibitem [{\citenamefont {Wang}\ \emph {et~al.}(2017)\citenamefont {Wang}, \citenamefont {Jacobson}, \citenamefont {Edwards},\ and\ \citenamefont {Clark}}]{10.21468/SciPostPhys.3.3.022}%
  \BibitemOpen
  \bibfield  {author} {\bibinfo {author} {\bibfnamefont {Y.-H.}\ \bibnamefont {Wang}}, \bibinfo {author} {\bibfnamefont {T.}~\bibnamefont {Jacobson}}, \bibinfo {author} {\bibfnamefont {M.}~\bibnamefont {Edwards}},\ and\ \bibinfo {author} {\bibfnamefont {C.~W.}\ \bibnamefont {Clark}},\ }\bibfield  {title} {\bibinfo {title} {{Induced density correlations in a sonic black hole condensate}},\ }\href {https://doi.org/10.21468/SciPostPhys.3.3.022} {\bibfield  {journal} {\bibinfo  {journal} {SciPost Phys.}\ }\textbf {\bibinfo {volume} {3}},\ \bibinfo {pages} {022} (\bibinfo {year} {2017})}\BibitemShut {NoStop}%
\bibitem [{\citenamefont {Butera}\ and\ \citenamefont {Carusotto}(2023)}]{PhysRevLett.130.241501}%
  \BibitemOpen
  \bibfield  {author} {\bibinfo {author} {\bibfnamefont {S.}~\bibnamefont {Butera}}\ and\ \bibinfo {author} {\bibfnamefont {I.}~\bibnamefont {Carusotto}},\ }\bibfield  {title} {\bibinfo {title} {Numerical studies of back reaction effects in an analog model of cosmological preheating},\ }\href {https://doi.org/10.1103/PhysRevLett.130.241501} {\bibfield  {journal} {\bibinfo  {journal} {Phys. Rev. Lett.}\ }\textbf {\bibinfo {volume} {130}},\ \bibinfo {pages} {241501} (\bibinfo {year} {2023})}\BibitemShut {NoStop}%
\bibitem [{\citenamefont {de~Nova}\ and\ \citenamefont {Sols}(2023)}]{PhysRevResearch.5.043282}%
  \BibitemOpen
  \bibfield  {author} {\bibinfo {author} {\bibfnamefont {J.~R. M.~n.}\ \bibnamefont {de~Nova}}\ and\ \bibinfo {author} {\bibfnamefont {F.}~\bibnamefont {Sols}},\ }\bibfield  {title} {\bibinfo {title} {Black-hole laser to bogoliubov-cherenkov-landau crossover: From nonlinear to linear quantum amplification},\ }\href {https://doi.org/10.1103/PhysRevResearch.5.043282} {\bibfield  {journal} {\bibinfo  {journal} {Phys. Rev. Res.}\ }\textbf {\bibinfo {volume} {5}},\ \bibinfo {pages} {043282} (\bibinfo {year} {2023})}\BibitemShut {NoStop}%
\end{thebibliography}%

\end{document}